\documentclass[fleqn,usenatbib]{mnras}
\usepackage{newtxtext,newtxmath}
\usepackage[T1]{fontenc}
\usepackage{ae,aecompl}
\usepackage{amsfonts} 
\usepackage{graphicx} 
\usepackage{subfig}
\usepackage{amsmath}	
\usepackage{amssymb}	
\usepackage{hyperref} 
\usepackage{gensymb}
\usepackage{color}

\title{The EDGE-CALIFA Survey: Exploring the Star Formation Law through Variable Selection}
\author[Dey et al.]{Biprateep Dey$^{1,2 \dagger}$, Erik Rosolowsky$^{2}$, Yixian Cao$^{3}$, Alberto Bolatto$^{4}$, 
\newauthor Sebastian F. Sanchez$^{5}$,  Dyas Utomo$^{6}$, Dario Colombo$^{7}$, Veselina Kalinova$^{7}$, 
\newauthor Tony Wong$^{3}$, Leo Blitz$^{8}$, Stuart Vogel$^{4}$, Jason Loeppky$^{9}$, 	Rub{\'e}n Garc{\'\i}a-Benito$^{10}$ \\
\\
$^{1}$School of Physical Sciences, National Institute of Science Education and Research, HBNI, Jatni-752050, India \\
$^{2}$Department of Physics, University of Alberta, 4-183 CCIS, Edmonton, Alberta, Canada \\
$^{3}$Department of Astronomy, University of Illinois, Urbana, IL 61801, USA \\
$^{4}$Department of Astronomy, University of Maryland, College Park, MD 20742, USA \\
$^{5}$Instituto de Astronom{\'i}a, Universidad Nacional Auton{\'o}ma de Mexico, A.P. 70-264, 04510 M\'exico, D.F., Mexico\\
$^{6}$Department of Astronomy, The Ohio State University, 140 West 18th Ave, Columbus, OH 43210, USA\\
$^{7}$Max Planck Institute for Radio Astronomy, Auf dem H\"ugel 69, D-53121, Bonn, Germany\\ 
$^{8}$Department of Astronomy, University of California, Berkeley, CA 94720, USA\\
$^{9}$Department of Statistics, University of British Columbia, Okanagan Campus, 3333 University Way, Kelowna, BC V1V 1V7, Canada\\
$^{10}$Instituto de Astrof{\'\i}sica de Andaluc{\'i}a, E-18008, Granada, Spain \\
$^{\dagger}$ \text{Present address:} Department of Physics and Astronomy, University of Pittsburgh, Pittsburgh, PA 15260, USA}

\date{}
\begin{document}
\maketitle
\pagenumbering{arabic}

\begin{abstract}
We present a multilinear analysis to determine the significant predictors of star formation in galaxies using the combined EDGE-CALIFA sample of galaxies.  We analyze 1845 kpc-scale lines of sight across 39 galaxies with molecular line emission measurements from EDGE combined with optical IFU data drawn from CALIFA.  We use the Least Absolute Shrinkage and Selection Operator (LASSO) to identify significant factors in predicting star formation rates.  We find that the local star formation rate surface density is increased by higher molecular gas surface densities and stellar surface densities.  In contrast, we see lower star formation rates in systems with older stellar populations, higher gas- and stellar-phase metallicities and larger galaxy masses.  We also find a significant increase in star formation rate with galactocentric radius normalized by the disk scale length, which suggests additional parameters regulating star formation rate not explored in this study.
\end{abstract}

\section{Introduction}
Star formation is one of the dominant secular evolution processes shaping galaxies over cosmic time. The  focus on how gas is converted into stars has guided the exploration of both nearby and distant galaxies \citep{kennicutt2012, madau2014}. Historically, the study of the ``star formation law'' has been through the conjecture by \citet{schmidtSFR} that the star formation rate is a function of the volume density of gas, establishing the local star formation rate through the free-fall time.  Empirically, the work of \citet{kennicutt1989, kennicutt1998} initiated the study of the star formation law in terms of the surface densities of (neutral) gas and the star formation rate, which represent the observationally tractable quantities in galaxies. Since local observations demonstrated that only the molecular phase of the interstellar medium (ISM) is associated with star formation, the star formation law studies began focusing on how the molecular gas content drives star formation \citep{wong2002, kennicutt2007}.  While these works found clear empirical relationships between star formation rate and gas content, though the extremes of the star formation environment, in low-metallicity systems and (U)LIRGs, show significantly different behaviours indicating that relationship defined in galaxy disks is not universal \citep{Daddi2010}.


Using a wealth of of multiwavelength data on nearby galaxies, it has become possible to explore the driving physics behind the star formation law, making quantitative tests of theories \citep{leroy2008, leroy2013, coldgass1, coldgass2}. In particular, two drivers of the star formation rate have appeared: the molecular gas content of the system \citep[e.g.,][]{leroy2013} and the depth of the stellar potential \citep[e.g.,][]{coldgass2}.  The contrast between these two studies is notable since they identify separate drivers for the star formation law, but they also employ different methods.  In particular, studies that find the local molecular gas content within galaxies drives the star formation rate typically favour line-of-sight based studies where kpc-scale regions of disks are being studied.  The studies favouring stellar content as the driving factor are usually drawn from studies of a wide range of galaxy types that focus on the global properties of galaxies.

Since star formation could be driven by multiple physical processes, several studies opt to consider several physical parameters simultaneously \citep{Dopita1994ApJ...430..163D, shi2011, leroy2013, shi2018}.  From these works, the focus on a pressure-based formalism \citep[e.g.][]{oml10} provides a means to synthesize the stellar potential and gas content contributions to the star formation law. Other models such as dynamical regulation \citep{silk1997, elmegreen1997} have also been explored empirically \citep{kennicutt1998, leroy2013}.  Recently \citet{colombo2018} also found that dynamical models could explain the empirical star formation law but that Hubble morphological type was a clearer factor in determining how the galaxies form stars. A similar conclusion was reached based on optical data alone \citep{GonzalezDelgado2014A&A...562A..47G, GonzalezDelgado2015A&A...581A.103G, GonzalezDelgado2016A&A...590A..44G}.

Thanks to dedicated campaigns, the number of multi-waveband data sets suitable for exploring galaxy evolution continues to grow.  In particular, the recent Extragalactic Database for Galaxy Evolution \citep[EDGE,][]{edge-califa} study combined a uniform sample of CO~(1-0) emission from 126 galaxies as observed with the Combined Array for Millimeter Astronomy (CARMA) with the integral-field unit spectroscopy made by the Calar Alto Legacy Integral Field Area survey \citep[CALIFA;][]{califa}.  The EDGE-CALIFA combined sample provides sufficient resolution to extend the line-of-sight based studies of molecular gas properties to a broader set of galaxy types while backing those data with the optical integral field unit (IFU) spectroscopy and modelling of the CALIFA survey.  The EDGE-CALIFA data set provides an excellent testbed for variations in the star formation law and has already been applied to explore this relationship, focusing on variations in depletion time in the centres of galaxies \citep{utomo2017} and the dynamical influences on star formation \citep{colombo2018}.

In this work, we explore the star formation law in the EDGE-CALIFA data from a different perspective, taking advantage of the rich database of galactic environments provided through the combined observational data. While the star formation law has been previously explored in the context of specific physical models, here we explore the data using machine learning methods designed to identify significant factors. This provides a data-driven selection of the important physical quantities, potentially offering new insights into the key factors shaping star formation in galaxies. Given the large number of observed quantities, we frame the question as one of {\it variable selection} where we seek to identify scalings that are the drivers of the star formation rate.  The primary challenge making this selection stems from many of the variables being physically linked. For example, nearly every property (gas content, stellar surface density, metallicity) scales with distance from the centre of the galaxy, leading to large covariance among the physical parameters.  Fortunately, the machine learning community has developed several methods for approaching such multi-covariant data sets and we rely on their established methods to identify quantities that are nonetheless significant. This approach, however, suffers from not having an underlying physical basis, but the outcomes can be used to identify the physical effects that could be considered in future explorations that aim to unravel the star formation law.  Our overall strategy is to determine how the star formation surface density ($\Sigma_{\mathrm{SFR}}$) and the molecular gas depletion time ($\tau_{\mathrm{dep}}\equiv \Sigma_{\mathrm{mol}}/\Sigma_{\mathrm{SFR}}$) depend on other measurements of the local galactic environment (e.g., stellar surface density, metallicity) and on the galaxy properties as a whole (e.g., morphological type).  This approach is similar to the fundamental plane analysis that is applied to early type galaxies, where we seek physical correlations in a high-dimensional space that are not apparent in bivariate analyses alone \citep[e.g.,][]{Bernardi2003AJ....125.1866B}.

In Section \ref{sec:data}, we outline the data used in this study and present the machine learning methods that guide our approach in Section \ref{sec:ML}.  We present the results of these analyses in Section \ref{sec:results}.

\section{Data}
\label{sec:data}

We use the data acquired as a part of the EDGE-CALIFA survey \citep{edge-califa}. The EDGE survey constitutes the largest interferometric CO survey of galaxies in the nearby Universe and provides CO maps of 126 galaxies. The survey data have an angular resolution of $\sim 4.5''$, which translates into a typical spatial resolution of  $\sim 1.4$ kpc given that the median distance to the galaxies is 64 Mpc.  The CO data were  obtained using the Combined Array for Millimeter-wave Astronomy (CARMA) interferometric array. These data were combined with the Integral Field Spectroscopy data obtained using the Calar Alto Legacy Integral Field Area (CALIFA) survey, obtained using the Calar Alto 3.5-m telescope, which provides an angular resolution of $\sim 2.5''$ \citep{CALIFA2015A&A...576A.135G}. Both the CALIFA and EDGE data are smoothed to a common angular resolution of $7''$. The maps are then sampled on a hexagonal grid with $3.5''$ spacing so the data are nearly independent. The combined data sets allow us to construct maps of gas, stellar metallicities, extinctions, extinction corrected star formation rates, stellar mass densities and age. In addition, we also have the global parameters of these galaxies obtained from the LEDA catalogue \citep{LEDA}.

The joint EDGE-CALIFA data set provides a huge number of possible variables for consideration in this analysis.  The CO data provide information on CO emission brightness, line width, and line-of-sight velocity.  The CALIFA data include both gas emission line and simple stellar population (SSP) analyses as provided by the {\sc Pipe3D} pipeline \citep[v2.2,][]{pipe3d, fit3d}. We use global galaxy data as tabulated in \citet{Walcher2014A&A...569A...1W}. From the possible set of variables we chose the set of resolved and global variables as shown in Table \ref{tab:varList} that could potentially be considered as key factors in controlling the star formation process. The variables under consideration are either {\it resolved} or {\it global}.  A resolved variable is measured on each line of sight through the galaxy and global variables hold for galaxies as a whole.  Variables can also be {\it continuous} or {\it discrete}, where continuous data can take any value and discrete data are fixed to a small number of values.  In particular, we include discrete variables to reflect categorical data such as whether the galaxy has a bar or not.  

For each variable under consideration, we examine the {\it variance inflation factor} (VIF), which measures the multi-collinearity with other variables in the data set. For example, many properties in galaxies change significantly with radius (star formation, stellar surface density, molecular gas surface density, metallicity, etc.).  Therefore, these variables are typically correlated with each other and the VIF is a measure of how this multi-collinearity will affect the results of the regression. The variance inflation factor of a variable $X$ is expressed as $\mathrm{VIF}_X=(1-R^2)^{-1}$ where $R^2$ is the coefficient of determination for a linear fit predicting $X$ given all the other independent variables.  Quantitatively, the VIF measures how much the uncertainty in the regression is increased by the presence of linear correlations between the predictor variables. If the variance inflation factor is high \citep[usually taken to be $>5$;][]{ISL}, the variable is multi-collinear with the other independent data and does not contribute new information to the analysis. The VIF analysis is particularly useful when compared to the standard covariance matrix analysis since it can identify correlations between more than two variables that limit the statistical power of the analysis. We retain all variables, even though some have large VIFs.  Instead, we rely on our regression method (see Section \ref{sec:ML}) to select which of these correlated variables is most relevant to the star formation law.

In framing our analysis, we also considered the mass-to-light ratio of the stellar population, but found it had a VIF significantly larger than all the factors considered here (15 vs $\sim 2$).  We removed this variable from our analysis because it was strongly correlated with several other factors linked to the age of the stellar population that had a clearer physical interpretation.  A simple model of the stellar population could be also be formulated using the mass-to-light ratio as a proxy for several other physical effects.


For quantities that depend on the surface brightness of emission, we deproject the quantities by multiplying by $\cos i$ where $i$ is the inclination of the galaxy. For the analysis, we transform most variables using a log-transform (base 10).  This reduces power-law scalings to a linear relationship, which is required since the variable selection techniques we will use are formulated in terms of linear models.  Categorical variables are treated as indicator variables, taking a value of 1 if the galaxy or line-of-sight is in the category and 0 otherwise.

\begin{table*}
\begin{tabular}{cccl}
\hline
Property & Type$^{1}$ & Units (before transforming) & Description \\
\hline
$\log(\Sigma_{\mathrm{SFR}})$ & C, R & $M_{\odot}~\mathrm{pc}^{-2}~\mathrm{yr}^{-1}$ & Log of star formation rate surface density \\
$\log(\Sigma_{\mathrm{mol}})$ & C, R & $M_{\odot}~\mathrm{pc}^{-2}$ & Log of molecular gas surface density \\
$\log(\sigma_{\mathrm{CO}})$ & C, R & $\mbox{ km s}^{-1}$ & Log of molecular gas velocity dispersion\\
$\log(V_\mathrm{rot})$ & C, G & $\mathrm{km~s}^{-1}$ & Maximum rotation speed estimated from molecular gas\\
$12+\log(\mathrm{O/H})$ & C, R & $\cdots$ & Gas-phase metallicities from Marino~et~al.(2013)\\
$\log(\Sigma_{\star})$ & C, R & $M_{\odot}~\mathrm{pc}^{-2}$ & Log of stellar surface density\\
$\log(\sigma_{\star})$ & C, R & $\mbox{ km s}^{-1}$ & Log of stellar velocity dispersion \\
$Z_{\star}$ & C, R & $\log(Z/Z_\odot)$ & Stellar metallicity relative to solar \\
$M_\mathrm{gal}$ & C, G & $M_\odot$ &  Stellar mass of galaxy \\
$\log(\tau_{\star,m})$ & C, R & yr & Mass-weighted stellar age from SSP analysis \\
$\log(\tau_{\star,l})$ & C, R & yr & Luminosity-weighted stellar age from SSP analysis \\
$A_V$ & C, R & mag & Implied $V$-band extinction estimated from across CALIFA bandpass\\
$\log(R/R_s)$ & C, R & kpc & Distance from the galaxy centre normalized by the scale length of the disk\\
$T$ & D, G & $\cdots$ & Numerical Hubble type \\
Bar & D, G & $\cdots$ & Galaxy is classified as having a bar \\
Ring & D, G & $\cdots$ & Galaxy is classified as having a ring \\
Multiple & D, G & $\cdots$ & Galaxy is classified as Multiple \\
Centre & D, R & $\cdots$ & Line-of-sight in centre of galaxy\\
\hline
\multicolumn{4}{l}{$^{1}$ C = continuous, D = discrete, R = resolved, G = global}
\end{tabular}
\caption{List of variables used for the multi-dimensional model}
\label{tab:varList}
\end{table*}

We measure the molecular gas surface density ($\Sigma_{\mathrm{mol}}$) from the EDGE data, using the CO integrated intensity values ($W_{\mathrm{CO}}$) calculated from \citet{edge-califa} and converting these into a molecular gas mass using a uniform CO-to-H$_2$ conversion factor $\Sigma_{\mathrm{mol}} = \alpha_{\mathrm{CO}} W_{\mathrm{CO}}$ with $\alpha_{\mathrm{CO}} = 4.35~M_{\odot}\mbox{ pc}^{-2}/(\mbox{K km s}^{-1})$ and $W_{\mathrm{CO}}$ being the integrated intensity of the CO emission \citep{Bolatto2013ARA&A..51..207B}.  The velocity dispersion of the molecular gas ($\sigma_{\mathrm{CO}}$) is calculated from the masked, emission-weighted second moment of the CO line.  We use the method described in Appendix B of \citet{Levy2018ApJ...860...92L} to correct $\sigma_\mathrm{CO}$ for the effects of beam smearing.  We further derive the peak rotation speed in the galaxy rotation curve ($V_\mathrm{rot}$) using the approach of \citet{Lelli2016AJ....152..157L} applied to the CO data.

We calculate star formation rates and gas-phase metallicities as per \citet{edge-califa} using the H$\alpha$ surface brightness and comparing the observed vs intrinsic H$\alpha$/H$\beta$ line ratio to infer the extinction $A_V$ assuming $R_V=3.1$. The extinction-corrected H$\alpha$ surface brightness is converted to a star formation surface density using the calibration of \citet{Kennicutt1998ARA&A..36..189K}:
\begin{equation}
    \frac{\mathrm{SFR}}{M_{\odot}~\mbox{yr}^{-1}} = 7.9\times 10^{-42}\left(\frac{L_{\mathrm{H}\alpha}}{\mbox{erg s}^{-1}}\right) 10^{{A_V}/2.5}.
\end{equation}
We use a local gas-phase metallicity measurement from the strong-line N2 indicator as calibrated in \citet{Marino2013A&A...559A.114M}. 

The parameters of the stellar environment are generated from the CALIFA data by Pipe3D \citep{fit3d, pipe3d}.  The Pipe3D package fits a set of simple stellar population templates to the stellar continuum of the galaxy, convolving the set of templates by the spectral response of the instrument and a line-of-sight stellar velocity dispersion ($\sigma_\star$) to identify an optimal match to the observed spectrum  \citep{GonzalezDelgado2015A&A...581A.103G}.  From the parameters of this template fitting, we determine the stellar surface density ($\Sigma_\star$), the stellar metallicity ($Z_\star$), the mass-weighted stellar age ($\tau_{\star,m}$), luminosity-weighted stellar age ($\tau_{\star,l}$), and the extinction to the stellar continuum across the CALIFA waveband (3745 \AA -- 7300 \AA), referencing the value to the $V$-band ($A_V$).

Finally, we use the CALIFA and LEDA catalogue values to determine the galaxy orientation parameters (inclination and position angle), the mass of the galaxy ($M_\mathrm{gal}$), how the galaxy is classified morphologically (numerical Hubble type $T$), and whether it hosts a bar, ring or is part of a multiple-galaxy system.  Finally, we use the orientation parameters to infer the mean galactocentric radius ($R$) for a given line of sight to and whether it is in the centre of the galaxy ($R<1$~kpc) or not.  The galaxies in the EDGE-CALIFA samples have different physical sizes, so we normalize $R$ by the exponential scale length of the stellar disk, $R_s$ derived in \citet{edge-califa}. 

The full sample in the EDGE-CALIFA overlap region contains $7800$ lines of sight at $7''$ resolution. We then apply a series of cuts to the data to ensure clean, significant detections throughout the sample.  First, we only include galaxies with inclination $i<70^{\circ}$ so that the deprojection correction is small (leaving 7001 data). We also require a joint detection of CO integrated intensity and H$\alpha$ emission at a 3$\sigma$ threshold (leaving 2074 data).  We removed a further 229 data using a series of criteria.  First, we also rejected all points that would be considered AGN above the \citet{Kauffmann2003} demarcation line based on a BPT diagram analysis \citep{Baldwin1981PASP...93....5B} as well as all points with an equivalent width in H$\alpha<6$~\AA\ since this emission is predominantly associated with older stellar populations \citep[$>0.5\mathrm{~Gyr}$,][]{Sanchez2014A&A...563A..49S}. After these down selections, we arrive at a final sample of 1845 independent data drawn from 39 different galaxies.  We tested whether there were any significant differences between the global galaxy parameters for the entire EDGE sample (126 galaxies) and those hosting lines of sight in this analysis using a two-sided Kolmogorov-Smirnov test.  We found no parameters were significantly different though the sample analyzed here is marginally closer (median distance is 56 Mpc vs.~68 Mpc) and less inclined than the overall EDGE population.

\section{Methods}
\label{sec:ML}

Galaxy-scale star formation is typically parameterized through the "Schmidt-Kennicutt law" or the star formation law \citep{schmidtSFR, kennicutt1998}.  This relationship uses a power-law scaling between the surface densities of star formation rate (SFR), $\Sigma_{\mathrm{SFR}}$ and molecular gas mass, $\Sigma_{\mathrm{mol}}$:

\begin{equation}
    \Sigma_{\mathrm{SFR}}=A \Sigma_{\mathrm{mol}}^{N} \label{sfrLaw};
\end{equation}
or with a logarithmic transform,
\begin{equation}
    \log(\Sigma_{\mathrm{SFR}})= \log A + N \log(\Sigma_{\mathrm{mol}}).
\end{equation}

\begin{figure}
    \centering
    \includegraphics[width=\columnwidth]{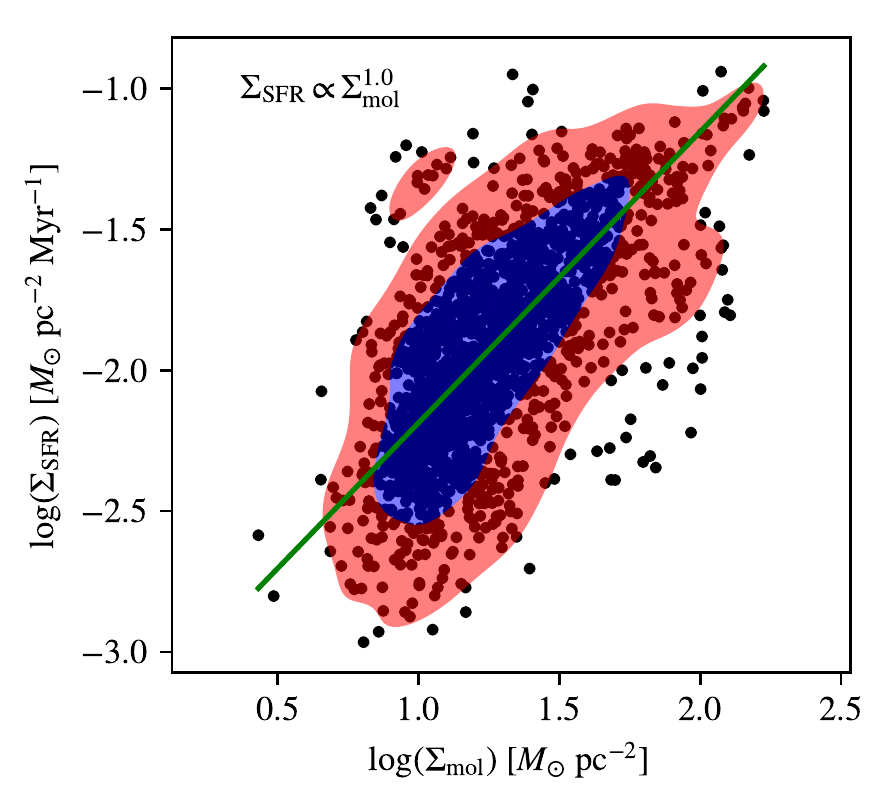}
    \caption{Star formation law drawn from our selected 1845 data out of the EDGE-CALIFA sample.  The solid line shows a robust regression between the data with $\Sigma_{\mathrm{SFR}}\propto \Sigma_{\mathrm{mol}}^{1.0\pm 0.1}$, consistent with other line-of-sight based work. The blue and red shaded regions contain 68\% and 95\% of the data respectively.}
    \label{fig:sfl}
\end{figure}

Our line-of-sight based analysis recovers the star formation law (Figure \ref{fig:sfl}).  We fit the star formation law using a linear regression in the logarithmic space.  Our sample data have significant internal scatter and do not have a well defined selection space in either of the axes (since the noise levels vary between the data sets). Thus, we treat the regression using a Theil-Sen estimator \citep{theil,theilsen} which determines the best fitting relationship as the median slope between the pairs of data in the sample.  This regression method is moderately robust to outliers and is efficient to calculate.  To further mitigate the effects of outliers, we perform random sample consensus \citep[RANSAC;][]{ransac} regression, which carries out repeated fits over subsets of the data to identify outliers and reduce the weight of these data in the fit.  RANSAC operates by drawing random subsamples of the data, performing the Theil-Sen regression within that subset, and then identifying outliers with respect to that regression.  Outliers for a subsample are identified as data are more than one median absolute deviation away from the trial regression line. The algorithm iterates, drawing new subsets, and identifying new outliers, until a stable set of outliers is identified, which are then ignored in the regression. The combination of these methods act to produce a robust fit in the presence of significant scatter and contamination by outlier data.  The star formation law in this analysis is $\Sigma_{\mathrm{SFR}} \propto \Sigma_{\mathrm{mol}}^{1.0\pm 0.1}$, consistent with similar approaches in the analysis \citep{leroy2013, edge-califa, utomo2017}. The outlier-rejection process identifies 30\% of the data as outliers. There is a scatter of 0.3 dex around the linear relationship. However, at low $\Sigma_\mathrm{mol}$ the star formation rates fall below the fit line, suggesting a non-linear form to the star formation law. Using RANSAC mitigates the influence of these data on the overall relationship. The deviation could reflect a real change in the star formation rate in this sample, potentially associated with other variables we will examine. These data are near the significance limit for our sample, and this the behaviour could reflect biases in our star formation rate estimators at these low rates. We proceed with a linear model in this work for consistency with previous work that considers the effects of completeness and censoring more carefully \citep{leroy2013} and return to discuss this choice in Section \ref{sec:caveats}.

When the index $N$ is consistent with unity, the data can be well described in terms of a near-constant molecular gas depletion time ($\tau_{\mathrm{dep}}$) with:
 \begin{equation}
     \tau_{\mathrm{dep}}\equiv \frac{\Sigma_{\mathrm{mol}}}{\Sigma_{\mathrm{SFR}}}.
 \end{equation}
The depletion time can be considered as the amount of time taken to convert all the gas present in the region into stars at the current star formation rate. It is found that the depletion time in galaxies is almost a constant with a typical value of 2.2 Gyr in nearby galaxies \citep{bigiel2008, leroy2013, utomo2017}. In our data set, we find a median depletion time of $\tau_{\mathrm{dep}} = 1.6\mbox{ Gyr}$.

\subsection{A Multilinear Star Formation Law}

We now consider the star formation law in terms of a multilinear fit where the star formation rate surface density, or alternatively the depletion time, are fit using a linear model.  We use log-transforms as summarized in Table \ref{tab:varList} to linearize the data.  The formalism of linear models is particularly well developed in the fields of statistics and machine learning, enabling this study to use relatively common methods. This analysis is completed with the {\sc Scikit-learn} \citep{scikit-learn} package in {\sc Python}.

 



We parameterize the star formation law using a linear model to a set of independent variables ($X_{ij}$ for the $i$th variable and $j$th datum) with a functional form:
\begin{equation}
    Y_j=C_0 + \sum_{i=1}^{p}C_{i}X_{ij}.
\end{equation}
In our exploration, the dependent variable $Y$ is either $\Sigma_{\mathrm{SFR}}$ or $\tau_\mathrm{dep}$. Typically, for unweighted fits, Ordinary Least Squares (OLS) regression is optimized by finding $C_i$ that minimize the residual sum of squares $S$ over $n$ data:
\begin{equation}
    S=\sum_{j=1}^{n}\left[Y_j -\left(C_0 + \sum_{i=1}^{p}C_{i}X_{ij}\right)\right]^2. \label{rss}
\end{equation}
However, OLS initially assumes that all variables are relevant explanatory variables for the data, but finding $C_i\approx 0$ would indicate that that factor is not relevant in predicting the dependent variable.  Since not all variables will necessarily be relevant for predicting $Y$, we can perform variable selection by including a regularization term in Equation \ref{rss}.  One approach is the Least Absolute Shrinkage and Selection Operator \citep[LASSO; ][]{lasso}.  The LASSO regression finds coefficients $C_i$ through a linear fit that minimizes the function:
\begin{equation}
\sum_{j=1}^{n}\left[Y_j -(C_0 + \sum_{i=1}^{p}C_{i}X_{ij})\right] ^2 + \alpha \sum_{i=1}^{p}|C_{i}|=S+\alpha \sum_{i=1}^{p}|C_{i}|
\label{eq:lasso}
\end{equation}
The extra term added is the LASSO penalty where the magnitude of the penalization is established by penalizing constant, $\alpha$.  The regularization term provides a penalty when a given coefficient ($C_i$) is non-zero, so the regression prefers a model with the minimum number of explanatory variables that predict $Y$.  Setting $\alpha=0$ returns to OLS and in the limit of $\alpha\to \infty$, $ C_i\to 0$. Intermediate values of $\alpha$ fixes some $C_i=0$ depending upon their significance as explanatory variables.  The formulation of this as the sum of the absolute values of the coefficients is important since it favours models where coefficients are exactly zero \citep{ISL}. Hence LASSO allows us to perform variable selection.

The parameter $\alpha$ needs to be determined for the regression. Since a single value of $\alpha$ applies to all variables, we standardize the data in the actual regression by subtracting off the mean values and normalizing each variable by its standard deviation.  This makes the numerical values of each axis carry uniform weight. We find the optimal value of $\alpha$ using cross-validation (CV) following the methods outlined in \citet[][their Section 2.5]{glmnet}. We first generate a grid of $10^3$ values of $\alpha$ ranging from the theoretical maximum ($\alpha_\mathrm{max}$) down to $10^{-3}\alpha_\mathrm{max}$, perform the regression for each of these $\alpha$ values, and assess which $\alpha$ value provides the most robust regression under CV testing.  For CV, we use $K$-fold CV with $K=5$, which entails partitioning the data into five random subsets.  Four of these subsets are used to establish the regression and the fifth is used as a test to determine how well the regression predicts their behaviour.  The test subset is then rotated among each of the groups while the process is repeated.  The optimal value of $\alpha$ is determined as the parameter that produces the most consistent fits between the training and validation sets.  After the regression is complete, we transform the resulting coefficients back to the scales that would be observed for the original data values instead of the standardized variables.

LASSO is well established for data without uncertainties, but the theoretical framework used to quantify the errors including data uncertainty is not complete \citep{ISL}. We therefore quantify the errors through a bootstrap resampling of the entire data set, repeating the regression including the full CV process on a subset of the data drawn with replacement from the originals.  The dependent variables are redistributed within their uncertainties with each bootstrap iteration.  We repeat the bootstrapping for 1000 iterations and report the uncertainties in the derived coefficients as the standard deviation of the coefficient values under the bootstrap. We note that we experimented with using a RANSAC approach to wrapping the entire LASSO process to mitigate the effects of outliers but we found that this did not produce a discernible change in our results.  By accounting for all the additional effects in the multilinear model, we find that the results are not significantly altered by outliers in our data.  Critically, the bootstrapping shows that the variable selection performed by LASSO is robust for nearly all variables: coefficients found to be zero remain zero for the vast majority of bootstrap iterations.

\subsection{Analysis of Mock Data}
\begin{figure}
    \centering
    \includegraphics[width=\columnwidth]{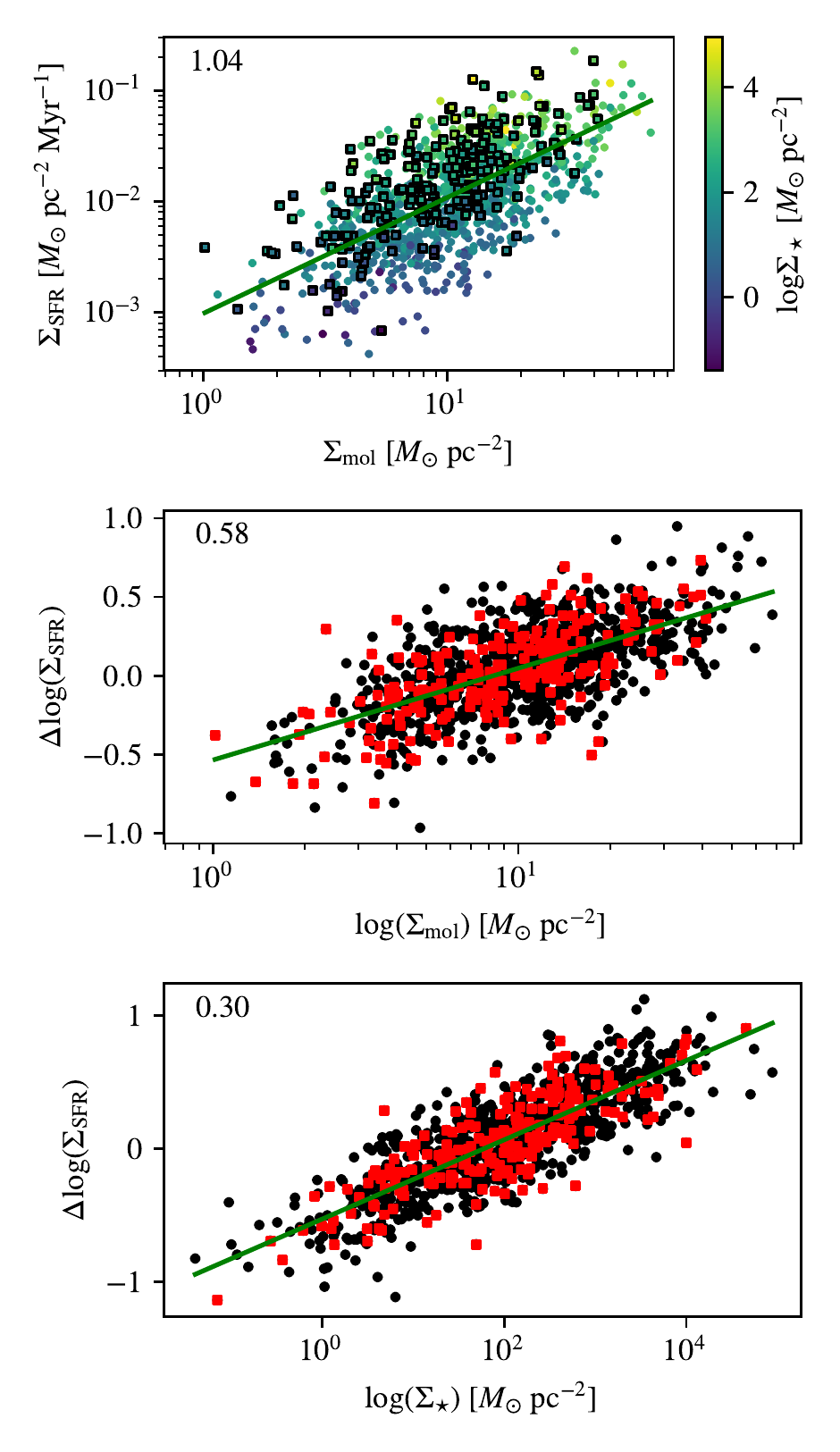}
    \caption{Mock data for a multi-linear star formation law.  The top panel shows the star formation rate plotted against molecular gas surface density, where the colour of each point indicates the log of the stellar surface density. The middle and bottom plots show partial regression plots for the molecular gas and stellar surface density respectively. The green line in the top panel show the RANSAC regressor applied to the $\Sigma_\mathrm{mol}$-$\Sigma_\mathrm{SFR}$ relationship.  In the bottom two panels, the green lines indicate the results of the LASSO fit to that variable.  Mock data assigned to barred galaxies are shown as square points in the panels and are coloured red in the bottom two panels.}
    \label{fig:mockdata}
\end{figure}

We demonstrate the LASSO approach to a multilinear star formation model by creating a mock data set consisting of 400 lines of sight.  We randomly generate molecular gas and stellar surface densities by drawing these values from correlated log-normal distributions.  The stellar and gas surface densities are assumed to have a correlation coefficient of $r=0.4$. Given these data, we create a mock star formation law 
\begin{equation}
    \Sigma_\mathrm{SFR} = 10^{C_0}\Sigma_\mathrm{mol}^a \Sigma_\star^b,
    \label{eq:mock}
\end{equation}
where we select $a=0.6$ and $b=0.3$ and $C_0$ is a constant.  We randomly assign 20\% of the the lines of sight into barred galaxies for which the star formation rate is increased by 0.2 dex.  Finally, we include 0.3 dex scatter in the SFR representing observational errors. We then process this mock data set through our analysis pipeline, mimicking the treatment that would be used in the real data.  In addition to the mock data, we also include an additional factor in the data drawn as random deviates from the standard normal distribution. This additional factor is not included in the SFR model (Equation \ref{eq:mock}), but it is included in the regression analysis, representing a variable that LASSO should set to zero. 

In Figure \ref{fig:mockdata} we show the results of the analysis.  The top panel illustrates the molecular gas star formation law.  The data have been coloured to illustrate the trend with stellar surface density and the barred galaxy sample is highlighted in square points.  Both higher stellar surface densities and being in barred systems are visibly at higher star formation rates, displaying the properties of our mock data.  The RANSAC regression to the molecular gas star formation law finds $\Sigma_\mathrm{SFR}\propto \Sigma_\mathrm{mol}^{1.0\pm 0.1}$, showing a shift from the expected value of $a=0.6$.  Performing the LASSO analysis on this model provides estimates of $a=0.58\pm 0.02$ and $b=0.30\pm 0.02$, consistent with the expected model values.  We also estimate that being in a barred galaxy increases the star formation rate by $0.22\pm 0.03$ dex and the additional random factor included in the model is correctly set to zero.

The correlation between the gas and stellar surface densities affects the results of the regression.  When these parameters are correlated, the slope of the gas-only star formation relationship is steeper than the intrinsic scaling with the molecular gas surface density.  With these mock data, we find $\Sigma_{\mathrm{SFR}} \propto \Sigma_\mathrm{mol}^{1.0\pm 0.1}$, like the observed data (Figure \ref{fig:sfl}). The multi-variate analysis is able to separate these effects, provided the correlation coefficient ($r$) is not near 1.0.  In exploring different mock data sets, we find that the model cannot distinguish between the influence of stars and molecular gas when $r>0.9$, but this threshold will vary with the sample. When we separate the effects of gas and stars, we identify shallower indices for these two parameters and their intrinsic correlation leads to a steeper index in the gas-only star formation law. The same conclusion has been reached by \citet{shi2011, shi2018}.  

To highlight the behaviour of the LASSO regression, we display a partial regression plot in the bottom two panels of Figure \ref{fig:mockdata}.  Here, we plot the $k$th variable against the residual of the star formation rate after correcting for all other variables in the fit:
\begin{equation}
\Delta Y_j = Y_j - C_0 -\sum_{i\ne k} C_i \left[X_{ij} - \langle X_{ij} \rangle\right].
\end{equation}
We subtract off the median of the data $\langle X_{ij}\rangle$ to account for the observed data not being centred on 0.  For example, the data shown in the middle panel of Figure \ref{fig:mockdata} is given by
\begin{equation}
    \Delta \log \Sigma_\mathrm{SFR} = \log \Sigma_\mathrm{SFR} -C_0 - 0.30 (\log \Sigma_\star - \langle \log \Sigma_\star\rangle) - 0.22 B,
\end{equation}
where $B$ is an indicator variable equal to 1 if the datum is in the barred sample in our mock data and 0 otherwise.  The scalars (0.30, 0.22) are from the regression to the mock data. This visualization subtracts off the effects of all the other factors in the model ($\Sigma_\star$, barred galaxy) to highlight the influence of the molecular gas on the star formation rate.  The slope of the line is $0.58$, which is the index of the molecular gas scaling alone. The approach performs well on these mock data, but the model is both simpler and has a cleaner data set than the actual observations.

\section{Results}
\label{sec:results}

We apply our variable selection methods to predict $\Sigma_{\mathrm{SFR}}$ and $\tau_\mathrm{dep}$ in terms of the variables given in Table \ref{tab:varList}.  We report the coefficients for these variables and their VIFs in Table \ref{tab:results} for linear models that explain $\Sigma_{\mathrm{SFR}}$ and $\tau_{\mathrm{dep}}$.  For modelling the depletion time, we do not include $\Sigma_{\mathrm{mol}}$ as an independent variable since it is in the depletion time expression.  Uncertainties in $\Sigma_{\mathrm{SFR}}$ are typically 0.07 dex given the flux errors in the H$\alpha$, H$\beta$ fluxes, but this does not include a potential systematic error due to changes in $R_V$.  The depletion time error includes uncertainties from both the CO and the Balmer lines and has a typical magnitude of 0.19 dex.  

\begin{table*}
    \centering
    \begin{tabular}{l r r r r r}
    \hline
    \multicolumn{1}{c}{Parameters} & \multicolumn{2}{c}{$\log(\Sigma_{\mathrm{SFR}})$} & &
    \multicolumn{2}{c}{$\log(\tau_{\mathrm{dep}})$} \\
    & VIF & $C_i$ & & VIF & $C_i$  \\
    \hline
$\log(\Sigma_{\mathrm{mol}})$ & 3.1 & $0.43 \pm 0.03$   &  & $\cdots$ & $\cdots$           \\
$\log(\sigma_{\mathrm{CO}})$  & 2.2 & $-0.21 \pm 0.06$  &  & 2.0      & $0.5 \pm 0.1$   \\
$\log(V_\mathrm{rot})$        & 4.1 & $-0.28 \pm 0.08$  &  & 4.1      & $0.3 \pm 0.1$    \\
$12+\log(\mathrm{O/H})$       & 1.8 & $-2.7 \pm 0.3$    &  & 1.8      & $2.6\pm 0.4$       \\
$\log(\Sigma_{\star})$        & 5.5 & $0.70 \pm 0.03$   &  & 3.0      & $-0.22 \pm 0.03$   \\
$\log(\sigma_{\star})$        & 2.9 & $0.19 \pm 0.02$   &  & 2.7      & $-0.33 \pm 0.03$   \\
$\log(Z_{\star}/Z_{\odot})$   & 2.6 & $-0.46\pm 0.08$   &  & 2.6      & $0.5 \pm 0.1$    \\
$\log(M_\mathrm{gal})$        & 2.8 & $-0.17\pm 0.02$   &  & 2.8      & $0.13 \pm 0.03$    \\
$\log(\tau_{\star,m})$        & 1.8 & $-0.56\pm 0.07$     &  & 1.7      & $0.4\pm 0.1$           \\
$\log(\tau_{\star,l})$        & 2.5 & $-0.12\pm 0.04$   &  & 2.4      & $0.00\pm 0.04$           \\
$A_V$                         & 1.8 & $0.00\pm 0.04$    &  & 1.8      & $0.00\pm 0.04$    \\
$\log(R/R_s)$                 & 2.8 & $0.58\pm 0.07$     &  & 2.8      & $-0.5\pm 0.1$           \\
$T$                           & 1.6 & $0.030 \pm 0.004$ &  & 1.7      & $-0.04 \pm 0.02$ \\
Bar                           & 2.2 & $-0.038 \pm 0.005$  &  & 2.2      & $0.037 \pm 0.005$    \\
Ring                          & 1.7 & $0.00 \pm 0.01$   &  & 1.7      & $0.00 \pm 0.01$   \\
Multiple                      & 1.1 & $0.00\pm 0.03$    &  & 1.1      & $-0.08 \pm 0.03$   \\ 
Centre                        & 1.4 & $0.11 \pm 0.03$   &  & 1.4      & $-0.09 \pm 0.04$    \\
    \hline 
    \end{tabular}
    \caption{Variance inflation factors (VIFs) and LASSO regression coefficients for different factors contributing to the star formation law. The VIFs indicate how correlated a given variable is with the other independent variables in the data set, with VIF $>5$ taken as a signature of significant correlation. The variable $C_i$ is the coefficient on the multilinear model and the uncertainties are established by resampling the data with the uncertainties and bootstrapping the sample.  Values of 0.00 indicate where the LASSO method has removed the variable from the model.}
    \label{tab:results}
\end{table*}

\begin{figure*}
\includegraphics[width=1\textwidth]{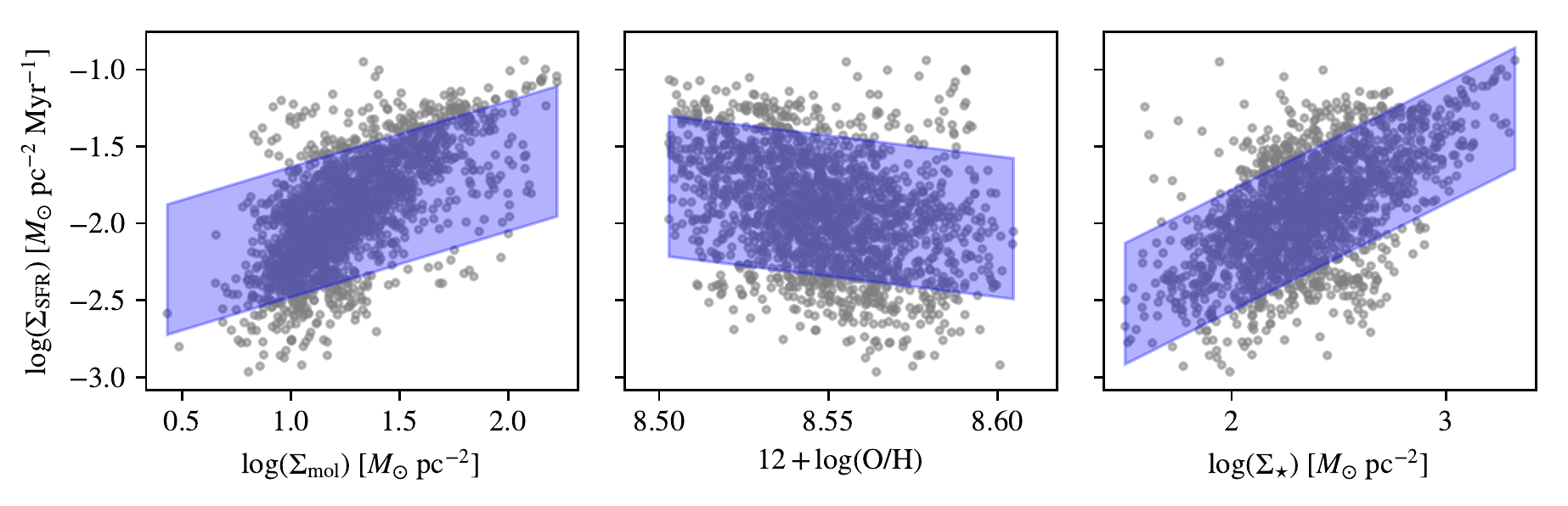}
\caption{Projection of multilinear model for the star formation rate ($\Sigma_{\mathrm{SFR}}$) into different bivariate spaces.  The 1845 data used in the fit are shown as grey points and the blue shaded region represents the projection of a portion of the hyperplane that runs through the full data set.  The correlation shows that increased molecular gas and stellar density corresponds to increases in star formation rate, but that high metallicity regions are associated with lower star formation rates.}
\centering
\label{fig:lassoSFR}
\end{figure*}
 
\begin{figure*}
    \centering
    \includegraphics[width=\textwidth]{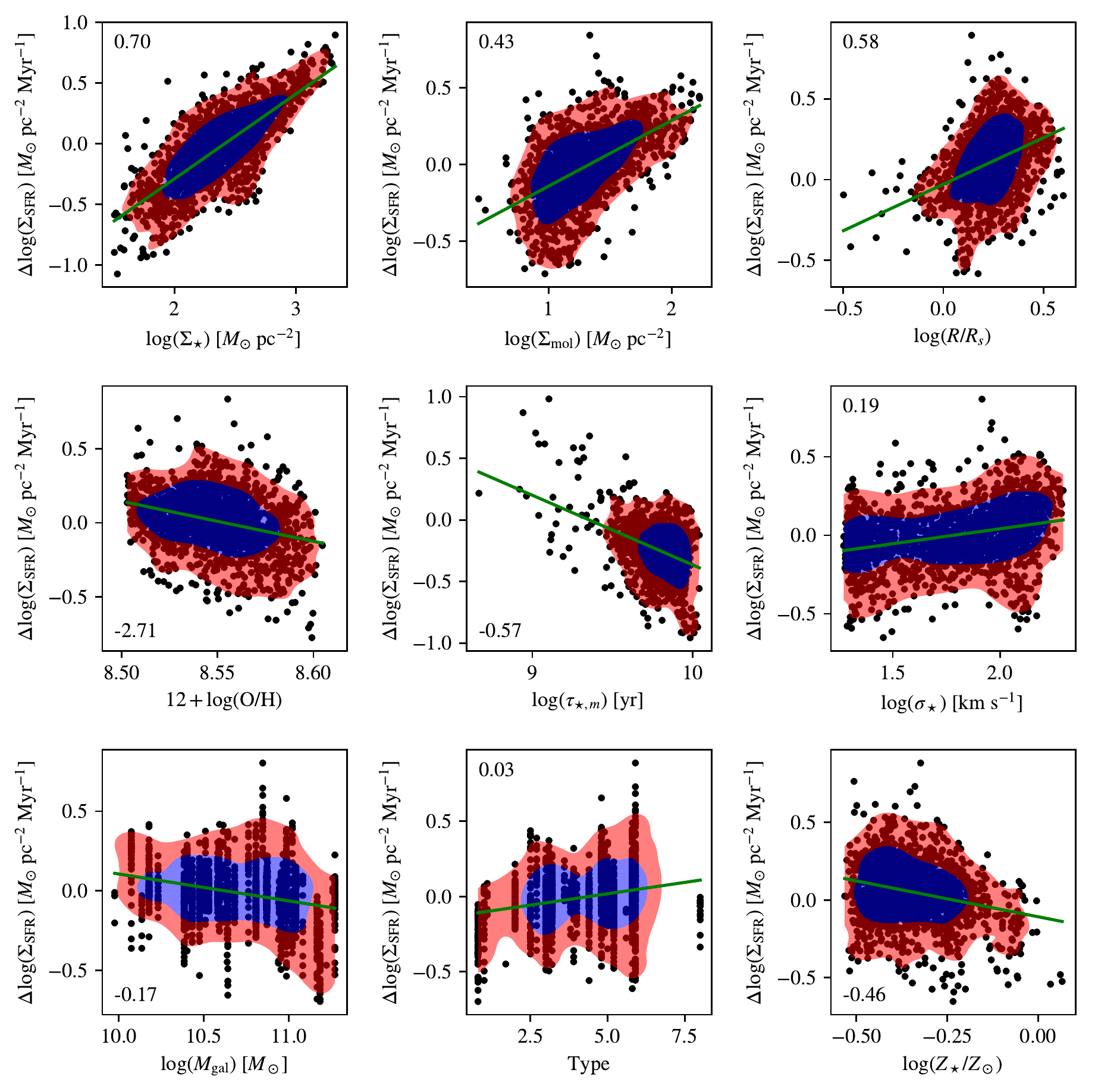}
    \caption{Partial regression for the nine most significant variables determining the star formation rate. The plot shows the residual star formation rate after subtracting off all contributions from factors except the factor plotted on the horizontal axis, which highlights the trends identified by the LASSO regression.  The green line indicates the best fitting multilinear relationship and the blue (red) contours contain 68\% (95\%) of the data.  The plots are ordered by the significance of the coefficients in the regression (left to right; top to bottom).  The numerical value for the slope of each partial regression is indicated in the corner of each plot.}
    \label{fig:SFLPR}
\end{figure*}

\begin{figure}
    \centering
    \includegraphics[width=\columnwidth]{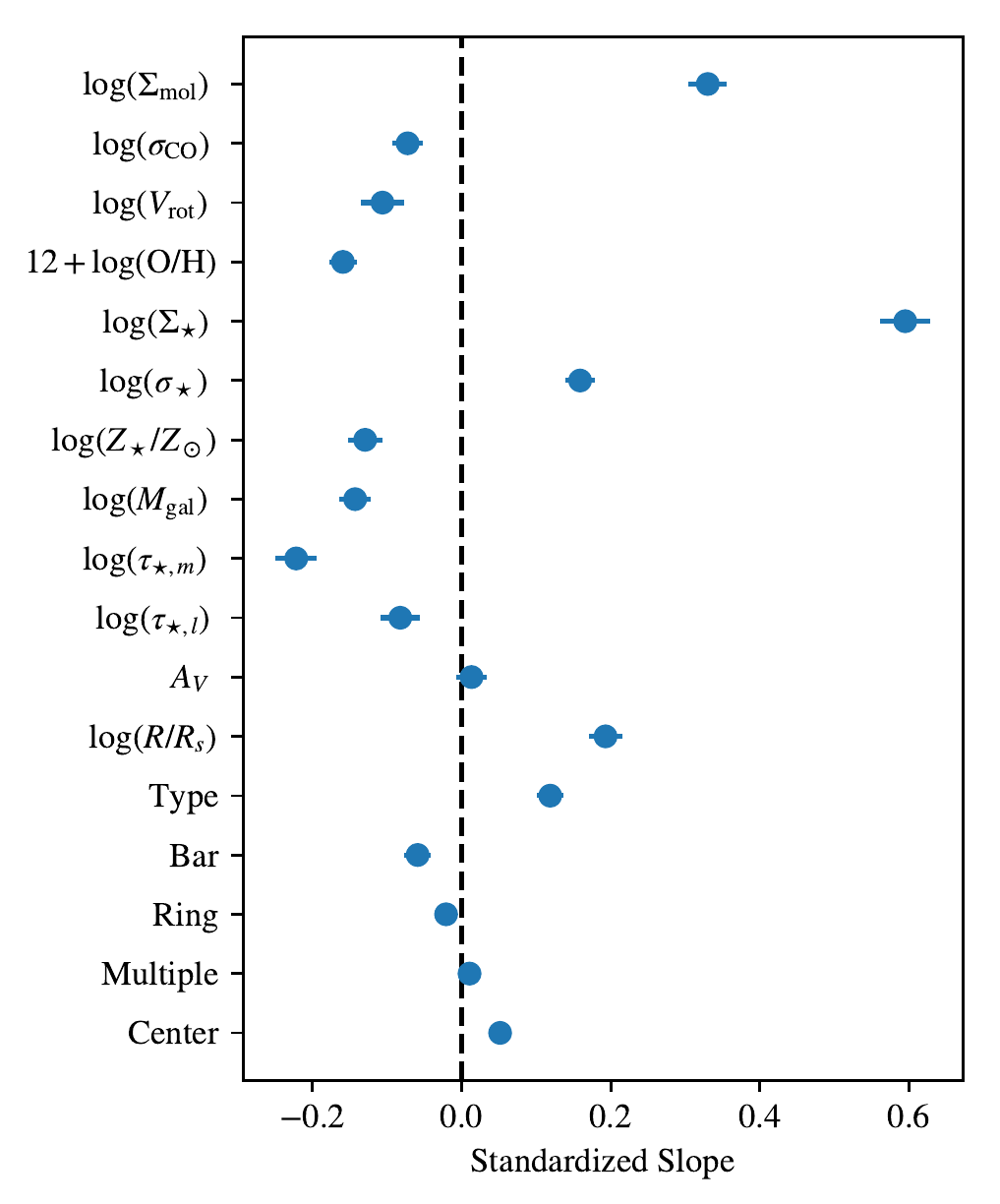}
    \caption{Standardized regression coefficients ($\widetilde{C}_i$) for contributing variables to the star formation rate. This plot shows the relative importance of different factors by scaling the slope of the relationship to the range of the data.  Error bars represent uncertainties in the fit parameters.  The most significant predictors of the star formation rate in this data set are $\Sigma_\mathrm{mol}$, $\Sigma_\star$, distance from the galactic centre, gas phase metallicity ($12+\log(\mathrm{O/H})$), stellar velocity dispersion, and age of the stellar population.}
    \label{fig:SC_SFL}
\end{figure}

\subsection{The Multilinear Star Formation Law}
\label{sec:mlsf}
In Figure \ref{fig:lassoSFR}, we show projections of the star formation law generated by the multilinear model for the selected variables we explored.  Because a multilinear model is a plane in the high dimensional parameter space, the shaded region shown in each of the fits represents the projection of that plane into the bivariate plot as shown.  The slope of each line is the coefficient $C_i$ given in Table \ref{tab:results} and the vertical extent of the region shows the possible values taken on by the multilinear regression based on the 5th to 95th percentile range of values seen in all the other dimensions.  We only show a few of the predictor variables, but overall the regression spans the general trends in the data, but there remain several outliers.  However, the regression is clearly different than the expectations for examining a single variable at a time (compare Figure \ref{fig:sfl} to the first panel of Figure \ref{fig:lassoSFR}). Showing the data in the original data-space gives a clear connection to the observations, but the regression hyperplane projects into the entire plotted region. The highlighted regime shows the portion of the hyperplane that would span the 90\% of the data closest to the plane.  

We also plot the partial regressions of the different factors against the incremental change in the star formation rate introduce that those factors in Figure \ref{fig:SFLPR}. These plots illustrate the significance of the different predictor variables and they are ordered by the significance of the coefficients from the regression.  The top, left panel of the Figure illustrates the regression against molecular gas, for which the tail to low star formation rates seen in Figure \ref{fig:sfl} is reduced but not eliminated.  The strong correlation with stellar surface density is also apparent in this analysis (top right).  We see clear declines in star formation with metallicity (gas and stellar), galaxy mass, and the age of the stellar population.  In particular, the decline in the star formation rate is particularly strong for the mass-weighted relationship, showing this effect is driven by the older stellar populations since these will contribute relatively more to the mass-weighting than to the luminosity-weighting.

The LASSO method is intended to identify factors that are irrelevant to the star formation law and several factors have low significance and are set to zero in a significant fraction of the bootstrap iterations.  This analysis finds that star formation rate can be explained without including the effects of the kpc-scale dust extinction for the stellar population ($A_V$) or the presence of a ring.  The lack of an effect from extinction suggests that the extinction correction for star formation rate based on the Balmer decrement is largely correct. These marginally significant factors include the CO velocity dispersion ($\sigma_\mathrm{CO}$), the rotation speed of the galaxy ($V_\mathrm{rot}$), the luminosity-weighted age of the stellar population ($\tau_{\star,l}$) and lines of sight through galaxy centres. The lack of a significant change based on the centres of galaxies arises because the sample mixes regions with enhanced star formation rates relative to their molecular gas content in with regions that have suppressed star formation \citep{utomo2017}.

If we consider only those factors that have coefficients larger that $8\times$ their uncertainty, this analysis argues for an empirical, multi-linear star formation law of the form:
\begin{equation}
    \Sigma_\mathrm{SFR} \propto \Sigma_\mathrm{mol}^{0.4} \Sigma_\star^{0.7} \left(\frac{R}{R_s}\right)^{0.6} \tau_{\star,m}^{-0.6} \sigma_\star^{0.2} Z_\mathrm{gas}^{-2.7},
    \label{eq:mlsfl}
\end{equation}
where we have defined $Z_\mathrm{gas}=10^{\log(\mathrm{O/H})}$.

The multi-linear star formation law highlights several important factors at setting the local star formation rate.  As expected, the molecular gas surface density is a clear, positive predictor of the SFR, but the model finds a relatively shallow slope for the scaling with $\log(\Sigma_{\mathrm{mol}})$: $0.43\pm 0.03$.  This shallow index contrasts with the gas-only index of the star formation law which shows a coefficient of $1.0\pm 0.1$ in our approach and others.  This linear model analysis favours a case where the other strong, positive predictor of star formation is stellar surface density ($\Sigma_\star$).  The stellar surface density traces the local gravitational field in disks and could be a strong controlling factor in star formation. The positive scaling with surface density is expected in models based on equilibrium density of the neutral medium being set by the self gravity of galactic disks \citep{oml10}, but the stellar velocity dispersion should have a suppressing effect if that model held in isolation since it would imply a higher stellar scale height and lower midplane stellar densities.

\begin{figure}
    \includegraphics[width=\columnwidth]{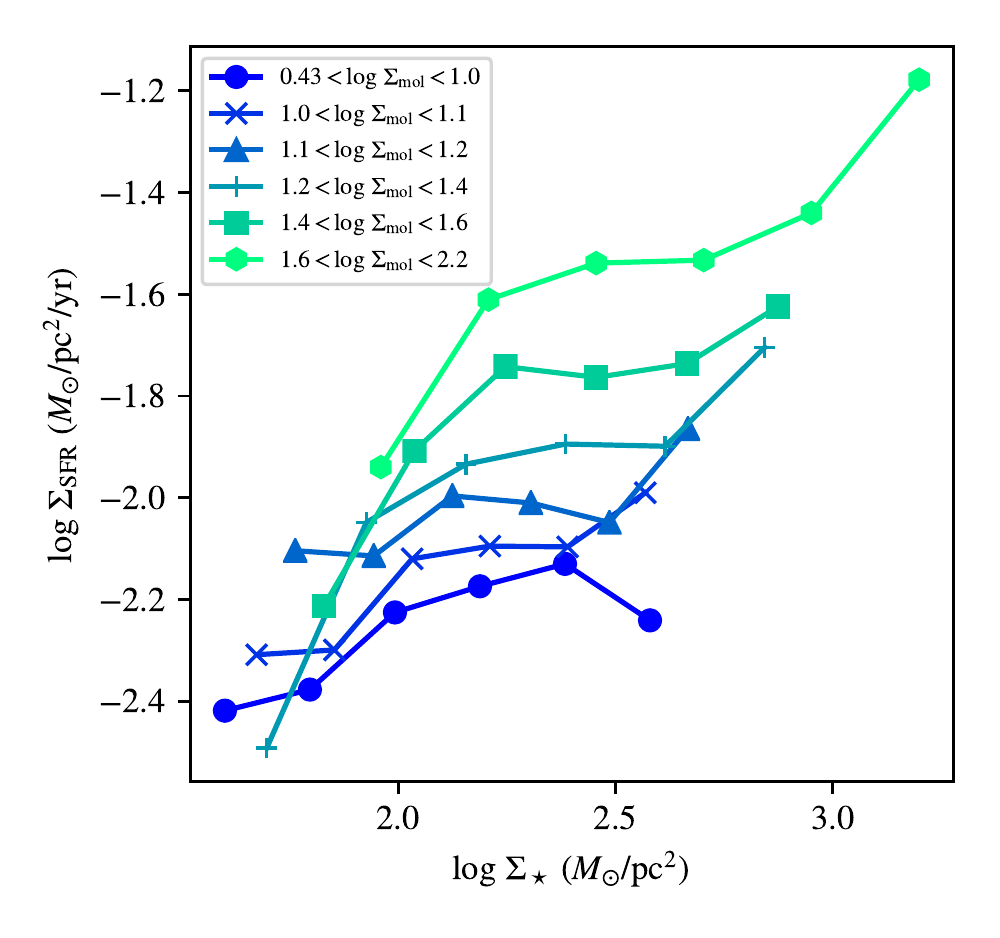}
    \caption{\label{fig:sfrbin} Scaling of the star formation rate ($\Sigma_\mathrm{SFR}$) as a function of stellar surface density ($\Sigma_\star$) for binned quantiles of the molecular gas surface density ($\Sigma_\mathrm{mol}$).  For each narrow range of the molecular gas surface density, the star formation rate shows a positive scaling with the local stellar surface density indicating the mutual influence of each of these parameters on the star formation rate.}
\end{figure}

The difference between a gas-only star formation law and a multifactor relationship can be reconciled because the molecular gas surface density and stellar surface density are correlated with each other. However, the application of this analysis suggests that the stellar surface density has a strong, independent influence on the local star formation that is more significant than the molecular gas content.  All lines of sight in our analysis, by construction, have detectable molecular gas content.  We further illustrate these two strong correlations in Figure \ref{fig:sfrbin} which shows the scaling between the star formation rate and the local stellar surface density in six equal quantiles of the molecular gas surface density.  Within these narrow ranges of the molecular gas surface density, the star formation rate shows a significant positive scaling with the local stellar surface density.  This figure illustrates that both of these factors (stellar and gas surface density) separately have positive correlation with the local star formation rate. Such a link with stellar population has been noted before \citep{edge-califa, shi2018} in single-factor analyses.  The binned trends in Figure \ref{fig:sfrbin} also show some curvature, indicating that a linear model is only an approximation to potentially more complex relationships between these data.  Of note, the scaling with stellar surface density appears shallower than the average value of 0.7 for galaxies of moderate surface densities.

Several factors are associated with a reduction in the local SFR, all of which are linked to more evolved galaxies.  In particular, larger galaxy masses, older stellar populations (as traced by the mass-weighted age), and earlier morphological types all suppress the local SFR. These can be broadly described as linked to older stellar populations, but the exact physical effects that yield the reduced star formation efficiency are unclear  \citep[see also ][]{colombo2018}.  With all other effects being held constant, molecular gas in a galaxy with an old stellar population forms stars less quickly compared to the same gas in a late-type galaxies with relatively young stars.  We also find that molecular gas found in barred galaxies has lower star formation rates by $-0.038\pm 0.005$ dex.

To display the relative importance of different factors in the model, we construct a {\it standardized slope} out of each coefficient $C_i$ and summarizes these values in Figure \ref{fig:SC_SFL}.  The standardized slope is determined from the standardized data used in the fit, i.e., the data scaled to have zero mean and a standard deviation of unity.  The standardized slope is related to the coefficients in Table \ref{tab:results} by
\begin{equation}
    \widetilde{C_i} = C_i \frac{\sigma_{X_i}}{\sigma_Y},
\end{equation}
where $\sigma_{X_i}$ and $\sigma_Y$ are the standard deviations of the corresponding variables.  The uncertainties in the coefficients are scaled by the same ratio and plotted as error bars.  This visualization shows the fractional change in the $Y$ variable (here, the star formation rate) predicted by the the change in the $i$th independent variable across the range of data contained in our analysis.  Points found close to zero indicate that the importance of this factor is relatively weak and the uncertainties indicate the significance of the estimate.

This visualization separates the dominant influences on the local star formation rate from lesser effects that nonetheless have a significant influence on the data.  The most significant effects here are encoded into Equation \ref{eq:mlsfl}, but several factors emerge as less important but nonetheless still significant.  The stellar metallicity and the luminosity-weighted mean age of the stellar population are both linked to reduction of SFR.  The mass-weighted age shows a more significant negative effect on the SFR, so whatever effect is measured here must be linked to the old stellar population.  

Similarly, lines of sight with higher gas-phase and stellar-phase metallicities also show relatively lower star formation rates.  Overall, the reduced levels of star formation associated with older stellar populations, earlier Hubble types, and higher rotation speeds may all be proxies for higher metallicities.  However, we do include both local stellar and gas phase metallicity as predictor variables and the LASSO method should eliminate these effects in favour of metallicities if this is the driving effect. 

If metallicity is a driving effect, the results we see may stem from our assumption of a constant CO-to-H$_2$ conversion factor underestimating the amount of molecular gas in these lower metallicity regions. This apparent suppression may emerge from a variable CO-to-H$_2$ conversion factor since the conversion factor will increase at lower metallicities.  This effect would imply more molecular gas than the CO measurements indicate in our model, which would imply a lower overall efficiency of the star formation process, making low- and high-metallicity systems more consistent in their star formation behaviour.  However, the conversion factor is thought to be relatively stable down to metallicities of $<0.5Z_{\odot}$ before increasing sharply \citep{Bolatto2013ARA&A..51..207B}.  Our data are typically found at higher metallicities, so this effect would probably be insufficient to explain the robustness of the anticorrelation of star formation rate with older stellar populations.  Even so, we have repeated our analysis using the variable CO-to-H$_2$ conversion factor prescription presented in \citep{Bolatto2013ARA&A..51..207B} using the gas-phase metallicity derived from the N2 indicator as an estimate of the local variation in metallicity.  Including these effects has no significant effect on this analysis.  The scaling coefficient for $\Sigma_{\mathrm{mol}}$ increases from $0.43\pm 0.03$ to $0.44\pm 0.03$.  Including the variable conversion factor marginally decreases the scaling of star formation rate with gas phase metallicity with the index changing rom $-0.27\pm 0.03$ to $-0.25\pm 0.03$ but there are no other significant changes in the derived star formation scalings. With variable conversion factor, the significant anticorrelation between older stellar populations and reduced star formation rates persists.  The variable conversion factor prescription also argues for a dependence on the surface densities of giant molecular clouds measured on $<10^2$ pc scales.  We are unable to assess these effects with our coarse resolution data, but nearly all of our lines of sight are through regions with surface densities $<10^2~M_\odot\mbox{ pc}^{-2}$ so these opacity effects are unlikely to change our results significantly.

The metallicity effect may be more subtle, where gas in high metallicity systems preferentially highlights ``sterile'' molecular gas that does not form stars.  In high metallicity systems, dust shielding of CO formation is more efficient, allowing CO to form at relatively low column densities.  This is the opposite effect from what is seen in low metallicity systems \citep{Bolatto2013ARA&A..51..207B}.  This low density gas could then be found outside of bound molecular clouds, which would manifest as a reduced star formation rate per unit molecular gas.

\begin{figure}
\includegraphics[width=0.5\textwidth]{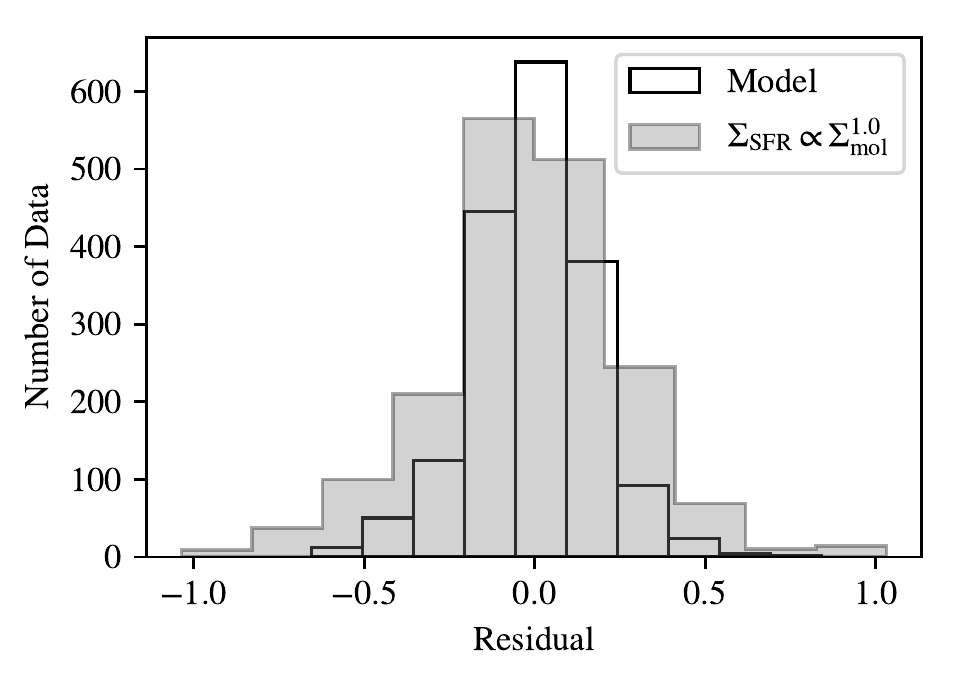}
\caption{Distribution of the residuals with respect to the gas-only star formation law and to the full LASSO-based linear model including all factors.  Including the extra factors in the star formation law reduces the scatter from 0.28 dex to 0.18 dex.  The scatter from the uncertainty in the star formation rate would be 0.07 dex, though we have not accounted for uncertainties in the independent variables.}
\centering
\label{fig:residualSFR}
\end{figure}

The significant positive scaling of star formation rate with normalized galactocentric distance likely indicates physical factors that have not been accounted for in our model but nonetheless influence the local star formation rate.  For example, several dynamical factors such as the epicyclic frequency or shear are not well represented in our sample data.  However, these effects have been hypothesized to play a major role in regulating the local star formation rate in galaxies \citep[e.g.,][]{Meidt2018ApJ...854..100M}.  Completing the analysis without normalizing by the disk scale length ($R_s$) eliminates the significance of the radial scaling but does not significantly alter the derived scalings for other parameters.  This suggests that the hidden parameters that are not being explored in this analysis are closely linked to the disk properties. 

A multilinear star formation law does reduce the scatter in the relationship compared to a gas-only star formation law.  In Figure \ref{fig:residualSFR}, we show the distribution of residual data for gas-only and the LASSO selected best-fitting relationship.  The intrinsic scatter drops from 0.28 dex to 0.18 dex.  The clearest systematic effect in the star formation law is the tail of points to lower star formation rates.  These are typically associated with the oldest stellar populations and earliest Hubble types present in our sample.  Including these parameters of the stellar populations noticeably reduces the tail to lower star formation rates, but does not eliminate its presence altogether. The scatter in the relationship remains in excess of that expected for the dependent variable ($\Sigma_{\mathrm{SFR}}$) but we do not account for uncertainties in the independent variables and several variables are discrete (Hubble type).  

\subsection{Multilinear Model for Depletion Time}


\begin{figure*}
\includegraphics[width=1\textwidth]{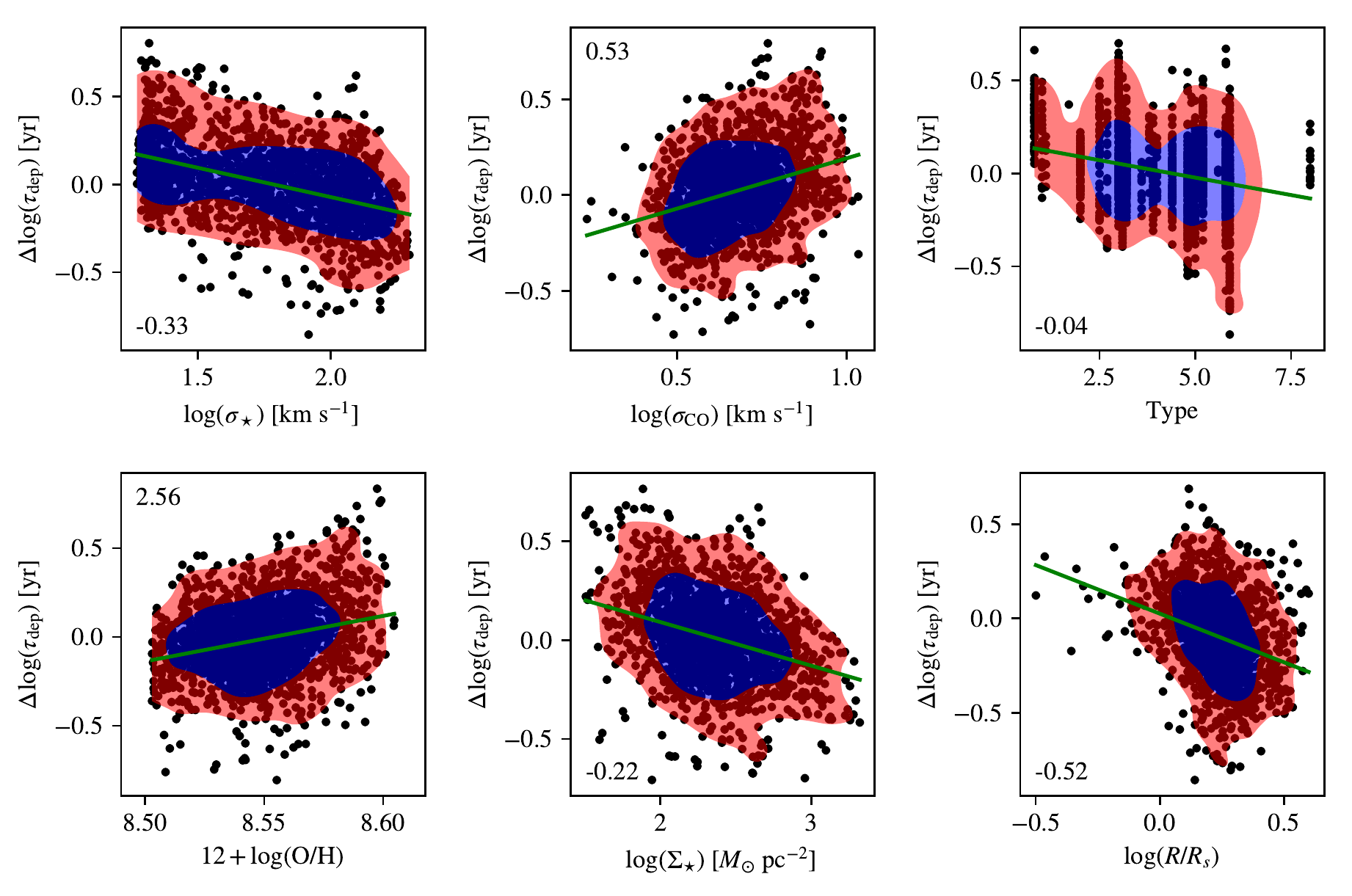}
\caption{Partial regression for the size most significant variables to the depletion time. The plot shows the residual depletion time after subtracting off all contributions from factors except the factor plotted on the horizontal axis.  The green line indicates the best fitting multilinear relationship and the blue (red) contours contain 68\% (95\%) of the data. Several factors independently affect the local depletion time. The plots are ordered by the significance of the coefficients in the regression (left to right; top to bottom).  The numerical value of the slope for the partial regression is indicated in the corner of each panel.}
\centering
\label{fig:TdepPR}
\end{figure*}

\begin{figure}
    \centering
    \includegraphics[width=\columnwidth]{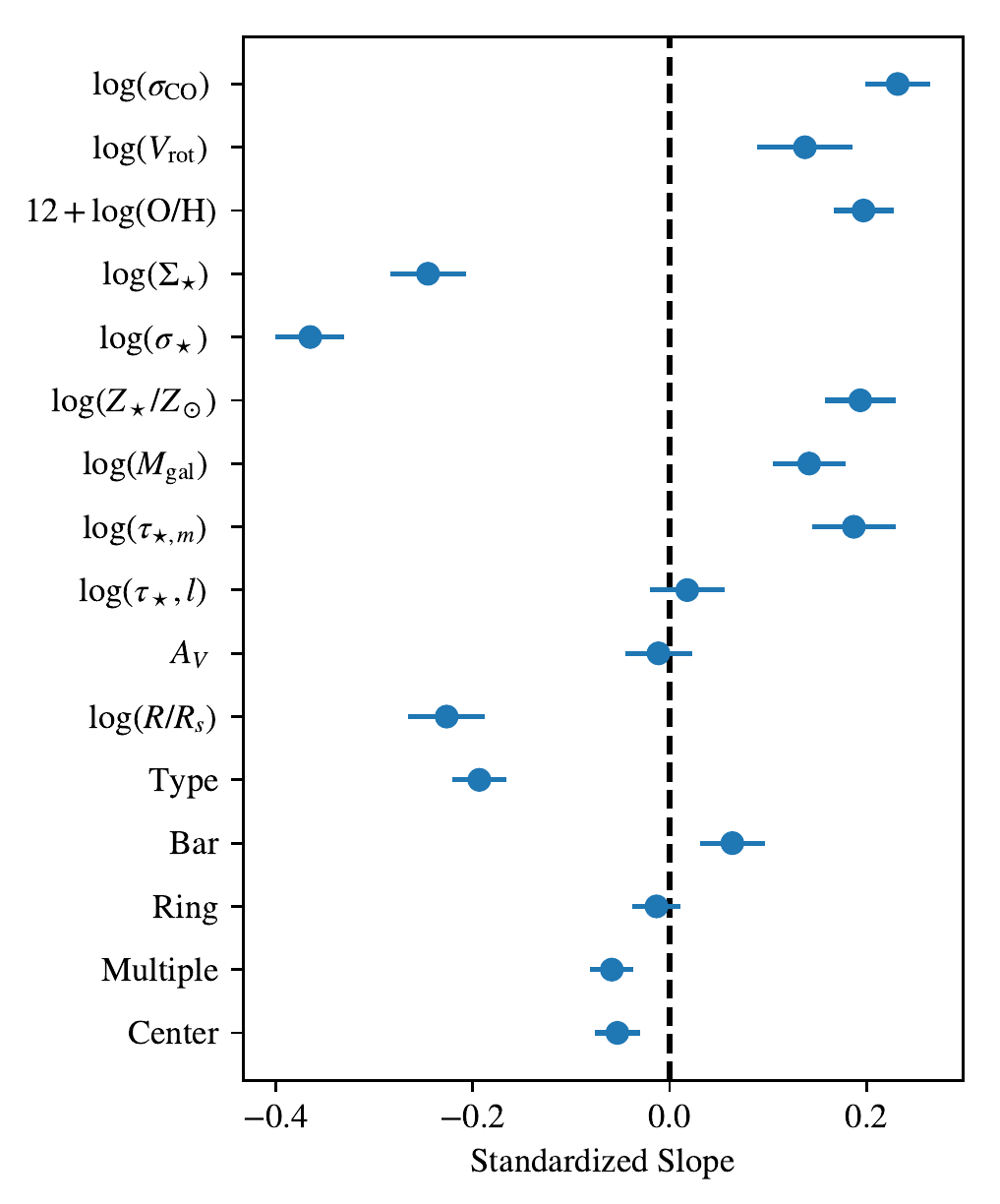}
    \caption{Standardized regression coefficients for different factors contributing to the molecular gas depletion time.  This plot shows the relative importance of different factors by scaling the slope of the relationship to the range of the data.  Error bars represent uncertainties in the fit parameters.  The significant predictors of the star formation rate in this data set are the velocity dispersions of gas and stars ($\sigma_\mathrm{CO}$, $\sigma_\star$), the stellar surface density ($\Sigma_\star$),  morphological type, gas phase metallicity, and distance from the galactic centre.\label{fig:SC_TDEP}}
\end{figure}


We also consider a model for the star formation in terms of the depletion time and compute a LASSO regression for the predictions of the depletion time in terms of all the parameters except for molecular gas surface density.  The results for the regression and the VIFs for the different factors are given in Table \ref{tab:results}. In Figure \ref{fig:TdepPR}, we show the partial regressions for several selected variables in the depletion time relationship and Figure \ref{fig:SC_TDEP} shows the scaled coefficients for  different factors.  

The regression leads to empirical model for the depletion time, based on scalings with coefficients $>6\times$ their associated uncertainties:
\begin{equation}
    \tau_\mathrm{dep} \propto  \sigma_\star^{-0.3} \sigma_\mathrm{CO}^{0.5} T^{-0.04} \Sigma_\star^{-0.2} Z_\mathrm{gas}^{2.6} \left(\frac{R}{R_s}\right)^{-0.5}
    \label{eq:mltdep}
\end{equation}
These results are consistent with the those seen for the star formation law except stellar velocity dispersion appears as a more significant predictor variable here, more so than stellar surface density, which is the opposite of what was found in the star formation law analysis.  We also find weak scalings in the depletion times which are longer for older stellar populations (higher $\tau_{\star,m}$), higher metallicities, higher galaxy masses, and earlier morphological types. Even though higher stellar surface densities and velocity dispersions are associated with older stellar populations, this analysis finds that these are separable effects that act in opposite directions.  

One notable difference with respect to the star formation law model is that higher values for the velocity dispersion of the molecular gas $\sigma_\mathrm{CO}$ predict longer depletion times.  There is a clear trend that larger $\sigma_\mathrm{CO}$ leads to longer depletion times, best seen in Figure \ref{fig:TdepPR}. This is generally consistent with expectations given the dynamical state of the gas. \citet{Leroy2017ApJ...846...71L} showed that the depletion times for molecular gas in M51 increased if the gravitational ``boundedness'' ($\propto \Sigma_\mathrm{mol}/\sigma_\mathrm{CO}$) decreased.  Our results act in the same sense, but the weak scaling in the star formation law approach (Section \ref{sec:mlsf}) shows that this effect may manifest from how we are framing the analysis.  We discuss this further in Section \ref{sec:caveats}.

\begin{figure}
\includegraphics[width=0.5\textwidth]{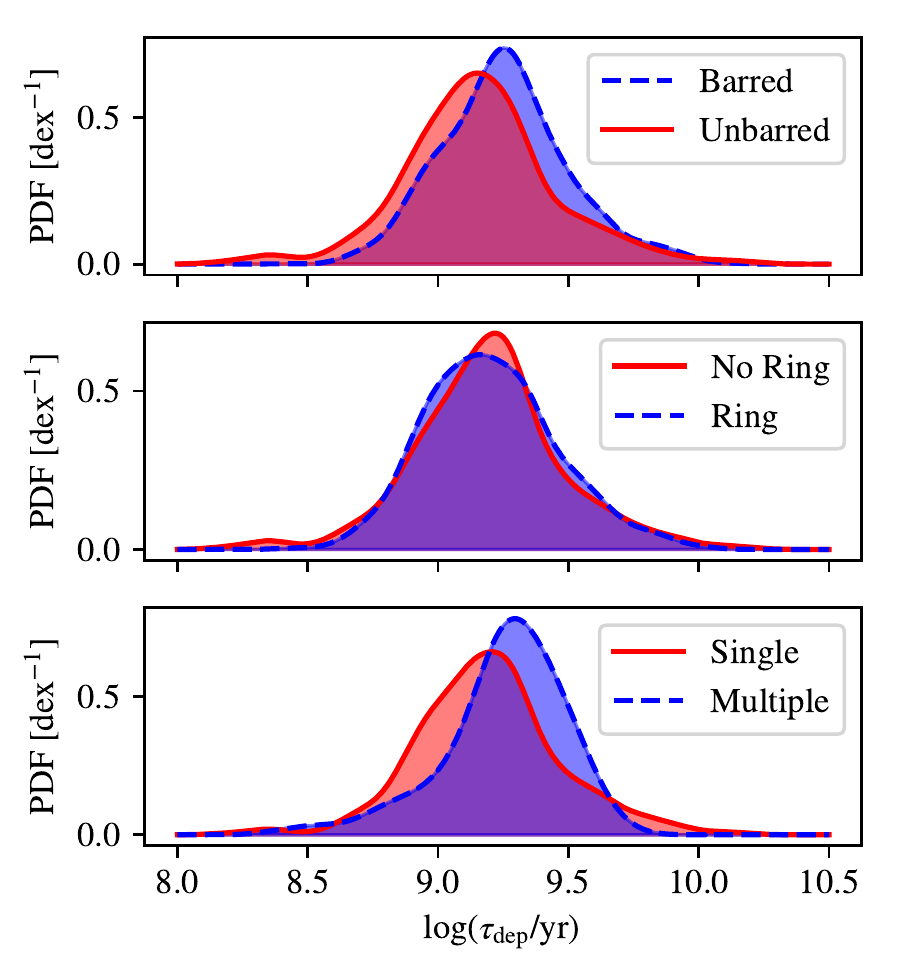}
\caption{Distributions of depletion times for systems with bars and rings.  Barred systems show slightly longer depletion times and rings do not have significant variation, which is visible in the probability density functions for the depletion time.  While the LASSO regression finds a weak trend that multiple galaxy systems have shorter depletion times, the appearance of the PDF (bottom panel) suggests the opposite.  This apparent inconsistency in the regression results from few points in a single system.}
\centering
\label{fig:violin}
\end{figure}

The morphological indicator variables (bar, ring, multiple) show more significant effects in terms of the depletion time model compared to the star formation law described in Section \ref{sec:mlsf}.  In Figure \ref{fig:violin}, we show the distributions of depletion times for the sample split into these different categories. These indicator variables are global: all lines of sight in a galaxy with a bar are considered as being in a barred galaxy, irrespective of whether the line of sight is actually located in the bar region.  Future work with resolved environmental descriptions will allow a more careful treatment of these effects.

As before, the linear regression indicates that the presence of a bar leads to a marginal reduction in the star formation rate and a corresponding increase in the depletion time but only by 0.04 dex.  We also noted the 0.04 dex decrease in the star formation rate found for the star formation law (Section \ref{sec:mlsf}).  The reduced star formation efficiency in barred galaxies initially appears at odds with the analysis carried out in \citet{Saintonge2012ApJ...758...73S}, which analyzed entire galaxy star formation efficiencies in the COLD GASS sample.  Like our analysis, they found that the effect of bars was not strongly distinguishable from the remainder of their sample.  The COLD GASS sample is a more representative selection from the galaxy population, but the EDGE-CALIFA sample is measuring the resolved star formation rate on smaller scales than in the COLD GASS sample.  Thus, our measurements are weighted on a by-area basis within galaxies whereas the COLD GASS sample is measuring overall efficiency of galaxies. While these morphological features may lead to significant changes in the local depletion time within galaxies, our analysis finds minimal effects from these factors on the galactic scale, especially when compared to the local environmental effects we are able to assess from this data set.

In both the star formation law and here with the depletion times, morphological classification as a ring galaxy does not lead to significant changes in the depletion time.  The multilinear depletion time model finds that galaxies classified as belonging to multiple systems have measurably shorter depletion times than the population as a whole. However, in Figure \ref{fig:violin}, most of the data flagged as coming from ``multiple'' systems have longer depletion times than the sample as a whole. However, there are only 85 of the 1845 of the lines of sight labelled as in multiple-galaxy systems drawn from five systems.  Most of the statistical weight in this result comes from the data drawn from the single galaxy NGC 3994 and excluding this system eliminates any difference in the depletion time from multiple-galaxy systems.  Examining the distribution of points in Figure \ref{fig:violin} shows that most lines of sight in multiple systems appear to have longer depletion times.  Thus, we do not consider this result to be significant.  In contrast, the ring and barred categories consist of 577 and 727 data respectively.

\begin{figure}
\includegraphics[width=0.5\textwidth]{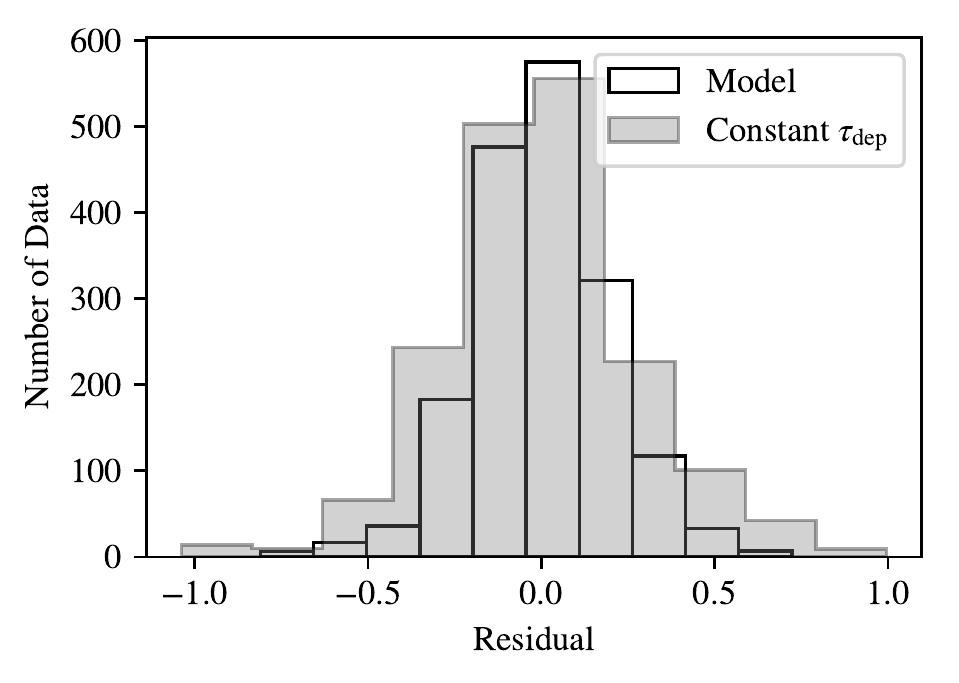}
\caption{Distribution of residuals for a constant $\tau_{\mathrm{dep}}$ model compared to the full LASSO-based linear model. Including extra factors in a linear model for the gas depletion time leads to a reduction of scatter from 0.28 dex to 0.20 dex.}
\centering
\label{fig:residualDT}
\end{figure}

In Figure \ref{fig:residualDT}, we show the residuals of the multilinear depletion time model compared to that of the constant depletion time model.  Again, there is a significant tail in the constant depletion time model and the multilinear model toward high depletion times.  The tail is somewhat reduced in the multilinear model, but the reduction in scatter is not as large for this model compared the star formation law.  The expected uncertainties in the observed depletion time estimate are 0.19 dex, and the reduction of 0.28 dex to 0.20 dex in scatter is approaching the limits of expected from noise.




\section{Discussion}
\label{sec:caveats}

This work focuses on using the tools of linear modelling and variable selection drawn from the field of machine learning and applied to the questions of the star formation law.  The initial results of this study are promising and connect many different lines of research seen individually through the literature.  However, there are clear areas where the work can be extended in future efforts.

\subsection{Context for Interpretation of Empirical Results}

Two key points emerge from this data-driven analysis of how galaxies form stars.  First, the data clearly show the gas-only star formation law has a linear scaling with $\Sigma_\mathrm{SFR} \propto \Sigma_\mathrm{mol}^{1.0}$; however, once additional factors are included, the scaling relationship with molecular gas becomes significantly shallower.  The star formation is usually considered to emerge from the properties of the cold molecular medium, as traced by the CO emission.  Here, we can consider star formation in the context of factors deriving from the galactic environment (i.e., everything but the properties of the cold ISM). Our analysis of mock data and the actual system argues that the intrinsic scaling with molecular gas surface density can be shallower than the nominal index of 1.0, with the correlations between different environmental factors, in particular stellar surface density, driving the resolved gas-only star formation law to the standard index of 1.0.

Taking our empirical models at face value (Equation \ref{eq:mlsfl} and \ref{eq:mltdep}), this would imply a strong environmental dependence setting the efficiency of the star formation process with the properties of molecular gas playing a secondary role.  Deeper explorations of the single-variable star formation law have also pointed to many of these same environmental effects to address the trends seen in earlier studies \citep{leroy2008, leroy2013, Meidt2013ApJ...779...45M, Meidt2018ApJ...854..100M, colombo2018}.  Instead, the analysis constantly identifies the role of the stellar environment as a key driver. Treated as a linear model, the stellar surface density is a positive predictor for the star formation rate.

The second key result that emerges is that we see evidence that the age of the stellar population anti-correlates with the star formation rate.  Moreover, the anticorrelation with the age of the stellar population is stronger for the mass-weighted age instead of the luminosity-weighted age. In Section \ref{sec:mlsf}, we argue that the effects of changing metallicity on the CO-to-H$_2$ conversion factor are unlikely to explain these effects as observational effects. We are faced with a significant and real effect: all else being equal, molecular gas found in environments with old stellar populations is less efficient at forming stars.

The significant dependence on age is likely a proxy for the dynamical state of galaxy: younger systems are dynamically cold (i.e., small random motions in the vertical and radial directions), and this enhances the star formation rate per unit molecular mass. Such effects could change the local stability of the stellar dynamical systems, but both the stellar surface density and velocity dispersion show up as positive predictors for star formation.  The velocity dispersions of the molecular gas and the stars are analyzed as the directly observed quantities. They have not been corrected into the components of the velocity ellipsoids for either of these species, though such analyses are becoming available \citep{Zhu2018MNRAS.473.3000Z}.  Further, this analysis considers kpc-scale resolution elements, so averaging over these large regions will also introduce significant beam smearing across coherent velocity structures.  As higher resolution or more careful analyses become available, a natural extension will be to include refined estimates of the dynamical environment in this type of analysis.

The influence of the local stellar population on the star formation rate also emerges from studies focusing on the optical IFU data alone
\citep[e.g.,][]{GonzalezDelgado2016A&A...590A..44G, Ibarra2016MNRAS.463.2799I, Garcia2017A&A...608A..27G, Belfiore2017MNRAS.466.2570B, Lopez2018A&A...615A..27L, Sanchez2018RMxAA..54..217S}.  The more detailed stellar analysis of these works make the case that the age of the stellar population is a good proxy for its current dynamical state, with older populations being dynamically hotter.  Moreover, these works also establish the morphological type correlations with the stellar population: earlier Hubble types have older, dynamically hotter stellar populations.  Optical IFU analyses can also infer the molecular gas content from extinction measurements \citep{Sanchez2018RMxAA..54..217S}, all of which find consitent analysis with the results presented here: molecular gas in older stellar populations is relatively less efficient at forming stars.  Star formation history analysis also suggests why systems with higher stellar metallicity have less efficient star formation: old systems had early bursts of star formaton, higher enrichment, and would be dynamically hotter today \citep{Vale2009MNRAS.396L..71V}. 
Physically, the likely origin of reduced star formation rate in dynamically hot systems arises from the increased stellar scale height for a given stellar surface density.  The molecular gas then follows the shallower stellar potential, reducing its average volume density and increasing its mean free-fall time \citep{oml10}.  

While consistent with the link to older stellar populations, this conjecture is undermined by our finding of a positive scaling with stellar velocity dispersion ($\sigma_\star$).  In isothermal stellar disks, the stellar volume density $\rho_\star \propto \Sigma_\star^2 /\sigma_{\star,z}^2$, where $\sigma_{\star,z}$ refers to the vertical velocity dispersion.  Thus, we would expect the scaling from the stellar velocity dispersion to have a negative coefficient where high vertical stellar velocity dispersions are associated with lower star formation rates.  Instead, we see the opposite: a weak positive correlation with $\sigma_\star$. Testing this conjecture will require a dynamical analysis to measure the vertical component of the stellar velocity ellipsoid, separate from the line-of-sight velocity dispersion on which our analysis is based.

The lower star formation rates seen in older stellar populations could also be a proxy for those systems where feedback from active galactic nuclei (AGN) quenches the star formation in these systems \citep{Springel2005ApJ...620L..79S}.  However, most quenching through feedback requires the AGN winds to drive material out of the galaxy.  This analysis highlights, instead, that the molecular gas already residing in such systems has a lower star formation rate.  Hence, our analysis shows that the effects of morphological quenching are visible in this sample \citep{Martig2009ApJ...707..250M}, though the analysis cannot quantify the relative importance of AGN-driven and morphological quenching. \citet{colombo2018} showed the clear scaling of depletion time with the morphological type of galaxies for systems with molecular gas, with earlier types have lower star formation rates.  However, \citet{Sanchez2018RMxAA..54..217S} also show that AGN are also preferentially found in morphologically earlier types of galaxies, so the effects may be difficult to distinguish.

We also see that the LASSO method identifies different factors as being significant when the analysis is recast in terms of the depletion time.  The choice to explore depletion time is generally motivated by the linear relationship between molecular gas and star formation rate surface density.  Since our multilinear star formation law shows a non-linear scaling, framing the analysis in terms of depletion time asks a different statistical question of the data.  Functionally, working with depletion times asserts that the amount of molecular gas is the primary factor at setting the star formation rate, so other effects are modifying a fundamental $\Sigma_\mathrm{SFR}\propto \Sigma_\mathrm{mol}^{1.0}$ relationship.  Asserting this scaling with index of 1.0 highlights other effects as important, notably the velocity dispersion of the molecular gas and the morphological (Hubble) types and shape classifications (bar, ring, multiple).  Like the analysis of \citet{colombo2018}, we find that the morphological type is a significant factor at regulating depletion time, of comparable significance and strength as the galaxy mass.  While the results for these two types of analyses are broadly consistent, the two approaches remain conceptually distinct when the scaling of star formation rate with molecular gas content is considered in a multi-factor analysis. New factors emerge in the depletion time relationship because the multilinear star formation law finds the index on the molecular gas scaling to be 0.4 not 1.0 (Equation \ref{eq:mlsfl}).  In a study of 15 nearby galaxies, \citet{Sun2018ApJ...860..172S} find a strong correlation between $\Sigma_\mathrm{mol}$ and $\sigma_\mathrm{CO}$ on 100 pc scales.  The stronger scaling with $\sigma_\mathrm{CO}$ manifests only the depletion time relationship because index of 0.4 in the star formation law is sufficient to account for the internal correlation between the molecular gas properties.

\subsection{Caveats and Future Work}

The clearest forthcoming improvement for this analysis is that we will be able to directly test the connection between the dynamical state of the stellar system and the star formation rate of the molecular gas at a given location.  This step will require full orbit decomposition of the galaxies, which is becoming possible for nearby systems \citep{Zhu2018MNRAS.473.3000Z}.  In particular, we will directly be able to test whether the stellar scale height explains all of the variation that manifests in our empirical star formation law in terms of Hubble type, stellar population age, and stellar velocity dispersion.

In carrying out the analysis, we tested the importance of the log of the distance to the target in the analysis, which does show up as a moderately significant effect in the star formation law ($C_D = -0.2 \pm 0.5$) but only results in small changes to the coefficients of the star formation law (e.g., $<1\sigma$ so the scaling for molecular gas shifts from 0.42 to 0.41).  However, this is an indicator that the EDGE sample, so analyzed, suffers some effects from Malmquist bias.  Specifically, the more distant galaxies are also those that are more luminous, with older stellar populations.  Ideally, the analysis would be executed at a common linear resolution.  However, the analysis already being carried out at kpc scales and reaching a common linear resolution for the whole sample would reduce many galaxies to single points, so future work would require higher resolution CO observations and deeper observations of more distant galaxies.

As the number of galaxies available for studies of the star formation law grows, it will become possible to examine the star formation law in terms of the galaxy cluster environment.  The initial studies of cluster effects on the star formation law have largely focused on binary tests \citep[cluster vs.~field galaxies, ][]{Vollmer2012A&A...543A..33V, Mok2016MNRAS.456.4384M}.  These studies find that cluster galaxies have longer depletion times compared to the field systems.  However, there are a range of different cluster environments, which readily could be integrated into a multilinear star formation model to assess their significance.

One other area for improvement in this work lies in the statistical formulation. The main advantage of this approach is that we can simultaneously assess all the possible factors contributing to the star formation rate.  However, to apply the LASSO method, we reduced the problem of star formation to a simple linear model.  While the regularization of the fitting through the $\alpha$ parameter (Equation \ref{eq:lasso}) is established through cross-validation, the linear model does not include either uncertainties in the independent variables, which are significant, nor does it include any contributions from the intrinsic variance of the data.  We attempt to include the the latter effect through the bootstrapping of the population.  We also have not fully modelled the censoring of our data based on the different selection effects present in the work.  Thus, the uncertainties for coefficients derived through the analysis (Table \ref{tab:results}) may be underestimated.  Finally, we adopted a linear model but there appears to be some non-linear behaviour in the full and partial regressions of star formation rate against molecular gas surface density.  For consistency with previous work, we used adopted a linear model to describe the star formation law.  Without a careful treatment of censoring and possible biases in CO or star formation tracers at low sensitivity, we cannot make strong claims about the shape of the star formation law. Given these limitations, a full Bayesian approach to the problem is merited in future work \citep{lasso, gelman2014bayesian}. However, the primary purpose of this work is to demonstrate the efficacy of variable selection approaches and identify the strongest positive and negative predictors of the star formation rate in galaxies through an empirical analysis.

\section{Summary}

Understanding what factors shape the star formation rate in galaxies drives a wealth of theoretical and observational work.  In this work, we have used the combination of CO(1-0) line observations of molecular gas with optical IFU data in the EDGE-CALIFA sample to explore a new data-driven approach to understanding what factors shape the star formation law.  Specifically, we have used the Least Absolute Shrinkage and Selection Operator (LASSO) modification to a linear model for the star formation rate to identify what variables most strongly predict the star formation rate surface density ($\Sigma_{\mathrm{SFR}}$) and local depletion time of the molecular gas ($\tau_{\mathrm{dep}}$).

Using the combined EDGE-CALIFA sample, we have identified a set of 1845 lines of sight spanning 39 galaxies that have well determined gas and stellar properties. Performing a simple linear model analysis on how the star formation rate scales with local molecular gas surface density recovers a similar star formation law as is found in other samples: $\Sigma_{\mathrm{SFR}} \propto \Sigma_{\mathrm{mol}}^{1.0\pm 0.1}$ \citep[cf.,][]{leroy2013,edge-califa,utomo2017}. Given the combined EDGE-CALIFA data sets, we identified a set of 18 possible factors (Table \ref{tab:varList}) that could contribute to the star formation rate. Given the LASSO model analysis, we find:

\begin{enumerate}
    \item The star formation rate increases with increasing molecular gas surface density, and stellar surface density.  The scaling with stellar properties is statistically more significant and stronger than the scaling with molecular gas surface density.
    \item While the star formation rate increases at higher stellar surface density, it also decreases for several factors associated with older stellar populations including the mass-weighted age of the stellar population, the gas- and stellar-phase metallicity and the mass of the galaxy. Earlier Hubble types are also linked to lower star formation rates.
    \item The star formation rate increases with galactocentric radius normalized by the disk scale length, suggesting additional paramters regulating the star formation rate not explored in this study.  One potential candidate for such a regulation mechanism is dynamical regulation of star formation through shear.
    \item Resolved star formation rates in barred galaxies are 0.04 dex lower than in non-barred galaxies after controlling for other factors, but this result is of marginal significance.

    \item The analysis of the molecular gas depletion time finds the same factors that the star formation law analysis does.  However, we also find depletion times are significantly longer in regions with higher molecular gas velocity dispersion.
    \item Developing a multilinear model for the star formation rate and the depletion times reduces the scatter in both of these relations, but neither model reaches the limits imposed by observational uncertainties in the dependent variables.
\end{enumerate}
With the advent of resolved, multiwaveband surveys, using a machine learning approach to analyses of the star formation rate will enable the identification of the key factors in galaxy evolution.  This approach is complementary to theoretical studies and provides a new way of assessing whether specific factors are actually important in different models.

\section*{Acknowledgments}

We are grateful to an anonymous referee whose review improved the clarity and robustness of the results.  BD is supported by a MITACS Globalink fellowship.  ER is supported by Discovery Grant from NSERC of Canada (RGPIN-2017-03987). DU is supported by the National Science Foundation under grants No. 1615105, 1615109, and 1653300. SFS thanks the support of the CONACYT grant CB-285080 and funding from the PAPIIT-DGAPA-IA101217 (UNAM) project. In addition to the software cited in the text, we gratefully acknowledge the use of {\sc matplotlib} \citep{matplotlib}, {\sc pandas} \citep{PANDASmckinney-proc-scipy-2010}, and {\sc astropy} \citep{Astropy2018arXiv180102634T, Astropy2013A&A...558A..33A}. RGB acknowledges financial support from the Spanish Ministry of Economy and Competitiveness through grant AYA2016-77846-P. We gratefully acknowledge conversations with Eric Koch and John Braun who suggested the LASSO approach to this data set. Support for CARMA construction was derived from the Gordon and Betty Moore Foundation, the Eileen and Kenneth Norris Foundation, the Caltech Associates, the states of California, Illinois, and Maryland, and the NSF. Funding for CARMA development and operations were supported by NSF and the CARMA partner universities. This study uses data provided by the Calar Alto Legacy Integral Field Area (CALIFA) survey (http://califa.caha.es/), based on observations collected at the Centro Astronómico Hispano Alem\'an (CAHA) at Calar Alto, operated jointly by the Max-Planck-Institut fűr Astronomie and the Instituto de Astrof\'sica de Andaluc\'a (CSIC).


\bibliographystyle{mnras}
\bibliography{bibliography}

\bsp    
\label{lastpage}
\end{document}